\documentclass[journal]{IEEEtran}
\usepackage[numbers,sort&compress]{natbib}
\usepackage{amsmath,amssymb,amsfonts,amsthm}
\usepackage{enumerate}
\usepackage{makecell}
\usepackage{algpseudocode}
\usepackage{algorithmicx,algorithm}
\usepackage{graphicx}
\usepackage{booktabs}
\usepackage[caption=false,font=footnotesize,labelfont=rm,textfont=rm]{subfig}
\usepackage{xcolor}
\usepackage{bm}
\usepackage{stfloats}
\usepackage{multirow}
\usepackage{cleveref}
\usepackage{empheq}
\usepackage[theorems,skins]{tcolorbox}
\usepackage{ragged2e}
\begin{document}
\newtheorem{proposition}{Proposition}
\newtheorem{lemma}{Lemma}
\bstctlcite{settingbib}

\title{Foundation Model Empowered Synesthesia of Machines (SoM): AI-native Intelligent Multi-Modal Sensing-Communication Integration}

\author{Xiang Cheng,~\IEEEmembership{Fellow,~IEEE}, Boxun Liu,~\IEEEmembership{Graduate Student Member,~IEEE}, \\Xuanyu Liu,~\IEEEmembership{Graduate Student Member,~IEEE}, Ensong Liu,~\IEEEmembership{Graduate Student Member,~IEEE}, \\and Ziwei Huang,~\IEEEmembership{Member,~IEEE}
\thanks{

Xiang Cheng, Boxun Liu, Xuanyu Liu, Ensong Liu, and Ziwei Huang are with the State Key Laboratory of Photonics and Communications, School of Electronics, Peking University, Beijing 100871, China (e-mail:xiangcheng@pku.edu.cn; boxunliu@stu.pku.edu.cn; xvanyvliu@gmail.com; ensongliu@pku.edu.cn; ziweihuang@pku.edu.cn).
}}

\markboth{}%
{Shell \MakeLowercase{\textit{et al.}}: A Sample Article Using IEEEtran.cls for IEEE Journals}
\maketitle
\begin{abstract}
To support future intelligent multifunctional sixth-generation (6G) wireless communication networks, Synesthesia of Machines (SoM) is proposed as a novel paradigm for artificial intelligence (AI)-native intelligent multi-modal sensing-communication integration.
However, existing SoM system designs rely on task-specific AI models and face challenges such as scarcity of massive high-quality datasets, constrained modeling capability, poor generalization, and limited universality. 
Recently, foundation models (FMs) have emerged as a new deep learning paradigm and have been preliminarily applied to SoM-related tasks, but a systematic design framework is still lacking.
In this paper, we for the first time present a systematic categorization of FMs for SoM system design, dividing them into general-purpose FMs, specifically large language models (LLMs), and SoM domain-specific FMs, referred to as wireless foundation models. 
Furthermore, we derive key characteristics of FMs in addressing existing challenges in SoM systems and propose two corresponding roadmaps, i.e., LLM-based and wireless foundation model-based design.
For each roadmap, we provide a framework containing key design steps as a guiding pipeline and several representative case studies of FM-empowered SoM system design.
{Specifically, we propose LLM-based path loss generation (LLM4PG) and scatterer generation (LLM4SG) schemes, and wireless channel foundation model (WiCo) for SoM mechanism exploration, LLM-based wireless multi-task SoM transceiver (LLM4WM) and wireless foundation model (WiFo) for SoM-enhanced transceiver design, and wireless cooperative perception foundation model (WiPo) for SoM-enhanced cooperative perception, demonstrating the significant superiority of FMs over task-specific models.}
Finally, we summarize and highlight potential directions for future research.

\end{abstract}


\section{Introduction}
\subsection{Background}
As a key infrastructure in the information age, the fifth-generation (5G) \cite{you2021towards} wireless communication networks have been widely deployed to support three major application scenarios, named enhanced mobile broadband (eMBB), massive machine type communications (mMTC), and ultra-reliable and low latency communications (uRLLC).
To meet the network demands beyond 2030, the sixth-generation (6G) wireless communication networks will further enhance and expand 5G to support a wide range of downstream applications, including cloud VR, Internet of Things (IoT) industry automation, cellular vehicle-to-everything (C-V2X), and digital twin.
As envisioned by the International Telecommunication Union (ITU) \cite{imt2021white} and other organizations, integrated sensing and communications (ISAC)  and integrated artificial intelligence(AI) and communications will play an unprecedentedly important role in achieving intelligent multifunctional wireless networks.

ISAC \cite{liu2022integrated} aims to unify radio-frequency (RF) sensing and communication functions in a single system, optimizing trade-offs and achieving mutual performance gains, referred to as RF-ISAC in this paper \cite{cheng2023intelligent}.
It is expected to improve spectral and energy efficiencies, reduce hardware and signaling costs, and foster deeper integration by co-designing communications and sensing for mutual benefits \cite{liu2023seventy}.
Nevertheless, the existing RF-ISAC is limited to RF sensing, failing to fully leverage communication and multi-modal sensing information. 
It is also restricted to static or low-speed scenarios \cite{cheng2023intelligent}, making it difficult to support high-dynamic 6G scenarios.
For instance, in typical 6G-enabled embodied intelligence scenarios \cite{ma2024survey}, a large number of intelligent agents move and interact in the complex and high-mobility environment while simultaneously deploying communication devices and sensors.
It is noted that each agent can capture communication information, as well as both RF data, e.g., radar point clouds, and non-RF sensing data, e.g., red-green-blue (RGB) pictures and light detection and ranging (LiDAR) point clouds.
However, most existing works, including RF-ISAC, focus on the separate study of communication and multi-modal sensory information, without considering their interconnections.
Therefore, there is an urgent need to systematically explore the intelligent integration and mutually beneficial mechanisms between communication and multi-modal sensing.

\subsection{Synesthesia of Machines}
\begin{figure*}[th]
    \centering
    \includegraphics[width=0.95\linewidth]{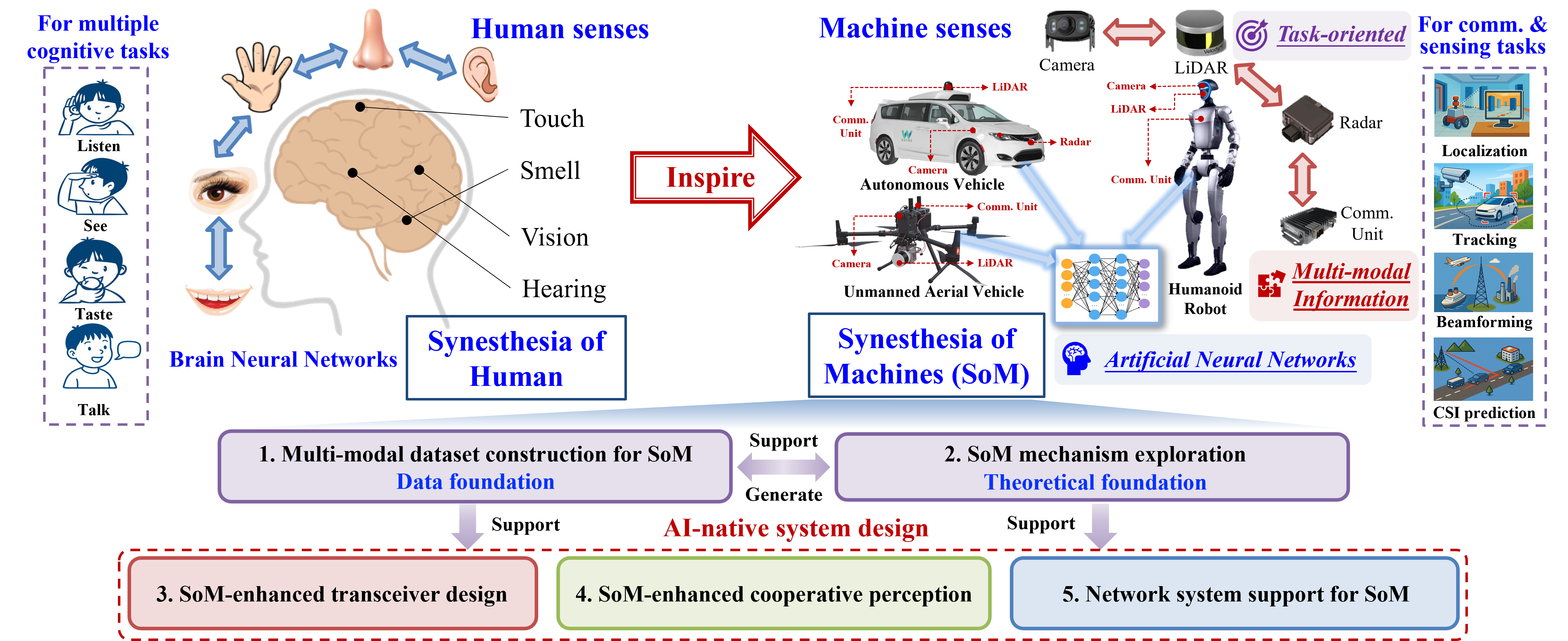}
    \vspace{-2mm}
    \caption{An illustration of the SoM framework, highlighting its three key characteristics and the interrelation of its five key research directions.}
    \vspace{-5mm}
    \label{SoM}
\end{figure*}
Inspired by the synesthesia of human, Synesthesia of Machines (SoM) \cite{cheng2023intelligent} has been proposed as a new paradigm for the intelligent integration of communication and multi-modal sensing.
Unlike RF-ISAC, SoM aims to enhance environmental sensing and communication functions through SoM processing of multi-modal data, including LiDAR point clouds, video, image, radar point clouds, and channel data.
However, SoM processing is not straightforward and faces significant challenges \cite{bai2024multi}.
First, there are significant differences in the representation of data across modalities. 
For instance, RGB images and depth maps provide dense two-dimensional(2D) visual information, while channel state information (CSI) consists of complex space-time-frequency dimensions.
Second, there are significant differences in the frequency bands used by different modalities, especially between non-RF sensing information and communication channel data, where the frequency bands differ by more than four orders of magnitude.
Third, there are differences in applications, where the objectives of multi-modal sensing tasks and communication tasks are inherently different.

As shown in Fig. \ref{SoM}, inspired by how synesthesia of human utilizes brain neural networks to process multi-sensory information for performing multiple cognitive tasks, the core of SoM processing lies in utilizing artificial neural networks (ANN) to handle multi-modal information for specific sensing and communication tasks.
Here, we summarize three key characteristics of SoM processing below.
\begin{itemize}
\item  \textit{Multi-modal information}: SoM processing fully leverages communication and multi-modal sensing information covering multiple frequency bands.
\item \textit{Task-oriented}: SoM processing focuses on specific sensing and communication tasks to design targeted algorithms.
\item \textit{Artificial neural networks}: SoM processing conducts task-oriented and data-driven neural network design.
\end{itemize}
In summary, SoM refers to task-oriented AI-native intelligence integration of communication and multi-modal sensing.
Specifically, to support the comprehensive design of SoM systems, five key research directions are outlined below, with their interrelationships illustrated in Fig. \ref{SoM}.
\begin{itemize}
\item \textit{Multi-modal dataset construction for SoM}: Since the scale and quality of the dataset determine the ultimate performance limit of AI-native systems, it is necessary to construct a massive and high-quality multi-modal sensing-communication dataset. {{Towards this objective, we construct a real-world data injected  synthetic intelligent multi-modal sensing-communication integration dataset, named SynthSoM, including RF communications, i.e., channel data, RF sensing, i.e., millimeter-wave (mmWave) radar data, and non-RF sensing, i.e., RGB images, depth maps, and LiDAR point clouds \cite{cheng2025synthsom}. The constructed SynthSoM dataset  provides a  reliable  data foundation for SoM research.}}

\item \textit{SoM mechanism exploration}: As the theoretical foundation of SoM research, the SoM mechanism, i.e., mapping relationship between communications and multi-modal sensing, needs to be explored based on the constructed multi-modal dataset. {The explored SoM mechanism not only supports the efficient and high-fidelity generation of multi-modal data, but also facilitates SoM-related research, including transceiver design,  cooperative perception, and network system support.}
However, due to significant differences between communications and multi-modal sensing in terms of data representation forms, acquisition frequencies, and application orientations, the SoM mechanism, i.e., mapping relationship, is complex and nonlinear, and thus is extremely difficult to explore.

\item \textit{SoM-enhanced transceiver design}: SoM-enhanced transceivers fully leverage rich prior knowledge from environmental multi-modal sensing information, such as the scatterer position and velocity, and further utilize the SoM mechanism to simplify or enhance the performance of wireless transmission systems.
The multi-modal dataset is the cornerstone of SoM-enhanced transceiver design, enabling AI-native multi-modal sensing-assisted wireless transmission systems.

\item \textit{SoM-enhanced cooperative perception}: SoM-enhanced cooperative perception focuses on improving perception performance under realistic communication constraints, such as quantization, multi-user interference, and channel effects.
Trained on multi-modal SoM datasets, the model incorporates physical-layer information to transmit multi-modal features that are well-suited to wireless channels, a process referred to as SoM feature transmission.
\item \textit{Network system support for SoM}: The collection, transmission, and processing of multi-modal sensing data rely on an integrated sensing, communication, computation, and storage network. A task-oriented network design supports complex SoM processing, which requires heterogeneous resources and is adaptive to available resources. Elastic task-oriented resource allocation schemes further improve resource utilization under a dynamic environment, while meeting the latency and availability service requirements.
\end{itemize}

However, existing research on SoM processing is still in its infancy and cannot adequately achieve the desired objective, attributed to the following four main challenges. 
\begin{itemize}
\item \textit{Scarcity of massive and high-quality datasets}: 
Considering that SoM processing is data-driven, high-quality datasets are the cornerstone of SoM processing. However, both real-world datasets constructed by measurement equipment and synthetic datasets constructed by simulation software encounter difficulties when constructing large-scale datasets.
Specifically, the real-world measurement method is limited by the high cost of multi-modal data collection devices, while the simulation method is constrained by the enormous computational cost with low quality of collected data.

\item \textit{Limited modeling capability}:
On one hand, SoM tasks are generally more challenging than conventional wireless system design, as they require modeling the complex mapping relationships between various modalities of information \cite{sun2024multi}. 
On the other hand, for scene-agnostic SoM tasks \cite{zhang2024integrated}, the scale of the training dataset is constrained by the high cost of real-world measurements. 
In such few-shot scenarios, task-specific deep learning-based schemes are impaired due to insufficient training data.

\item \textit{Limited generalization across data}:
Most existing ANNs used for SoM processing are trained and tested on specific datasets, exhibiting poor generalization. Specifically, when the data distribution undergoes significant changes, the performance of the ANN drops significantly. In this case, fine-tuning in new scenarios incurs additional data collection and network training overhead.

\item \textit{Limited universality across tasks}:
{Existing SoM processing schemes design distinct ANNs and loss functions for each specific task, thereby restricting the model to single-task learning.}
However, in real-world application systems, intelligent agents are required to handle a wide variety of SoM tasks.
Therefore, numerous separate ANNs need to be deployed simultaneously, resulting in significant model storage and management overhead.
\end{itemize}

\vspace{-2mm}
\subsection{Foundation Models}
In the past decade, deep learning has been widely applied across various fields, including communication and multi-modal sensing tasks, due to its powerful modeling ability directly from raw data without relying on prior information.
Nevertheless, conventional deep learning networks, referred to as \textit{task-specific models} in this paper, are trained on specific tasks and datasets in a supervised manner, generally lacking generalization and universality.
Recently, foundation models (FMs) \cite{bommasani2021opportunities} have emerged as a new paradigm in deep learning, revolutionizing several fields like natural language processing (NLP) and computer vision (CV).
Specifically, a foundation model \cite{bommasani2021opportunities} is any model that is trained on massive data, generally in a self-supervision manner, that can be adapted to a wide range of downstream tasks with fine-tuning or zero-shot inference.
Its core idea lies in \textit{pre-training and task adaptation}, i.e., during the pre-training phase, the model learns generalizable representations, and during deployment, it quickly adapts to specific scenarios with minimal or no data.
Large language models (LLMs) \cite{zhao2023survey} are the forefront and most successful representative achievements of FMs, where GPT-4 and DeepSeek \cite{liu2024deepseek} have validated the emergence of astonishing understanding and reasoning capabilities of FMs, driven by the enormous scale of model parameters and datasets.
In addition to general-purpose FMs for NLP or CV, domain-specific FMs are constantly emerging for various fields, such as time series prediction \cite{liang2024foundation}, weather forecasting \cite{bi2023accurate}, and remote sensing \cite{sun2022ringmo}.
Inspired by the powerful inference and generalization capabilities of FMs, we wonder \textit{whether FMs can be leveraged for SoM system design to address the aforementioned four challenges}.

\subsection{Related Works}
{Although research on FM-empowered SoM system design is still in its infancy, a growing number of studies \cite{xu2024large,qin2025generative,jiang2024large1} have explored the application of FMs in SoM-related fields, including dataset generation, wireless communications, RF-ISAC, and semantic communications.}
According to the type of adopted FMs, we classify these studies into two categories, including those based on general-purpose FMs, specifically LLMs, and those based on domain-specific FMs, namely \textit{wireless foundation models} as defined in Section \ref{section 2}.
Existing surveys on FM-empowered SoM-related fields have solely summarized a limited subset of such studies from narrow perspectives.
{For instance, survey \cite{cheng2023intelligent} first proposed and elaborated on the SoM concept and framework in detail, while its design approaches still mainly rely on task-specific deep learning models, with limited exploration of FM-enabled designs.}
Survey \cite{zhou2024large} provided a comprehensive overview of LLM-enabled telecommunication (telecom) networks, covering fundamental LLM techniques, key telecom applications, and future directions. 
The authors in \cite{zhou2024large} preliminarily explore the concept of domain-specific FMs for wireless prediction tasks, but do not incorporate the latest emerging research on wireless foundation models.
The review \cite{long2024llms} summarizes existing research on LLM-driven synthetic data generation, curation, and evaluation, but ignores studies on dataset generation for communications and multi-modal sensing.
In summary, a comprehensive and unified framework for FM-empowered SoM system design is still lacking in the existing literature.
\begin{figure*}[th]
    \centering
    \includegraphics[width=0.8\linewidth]{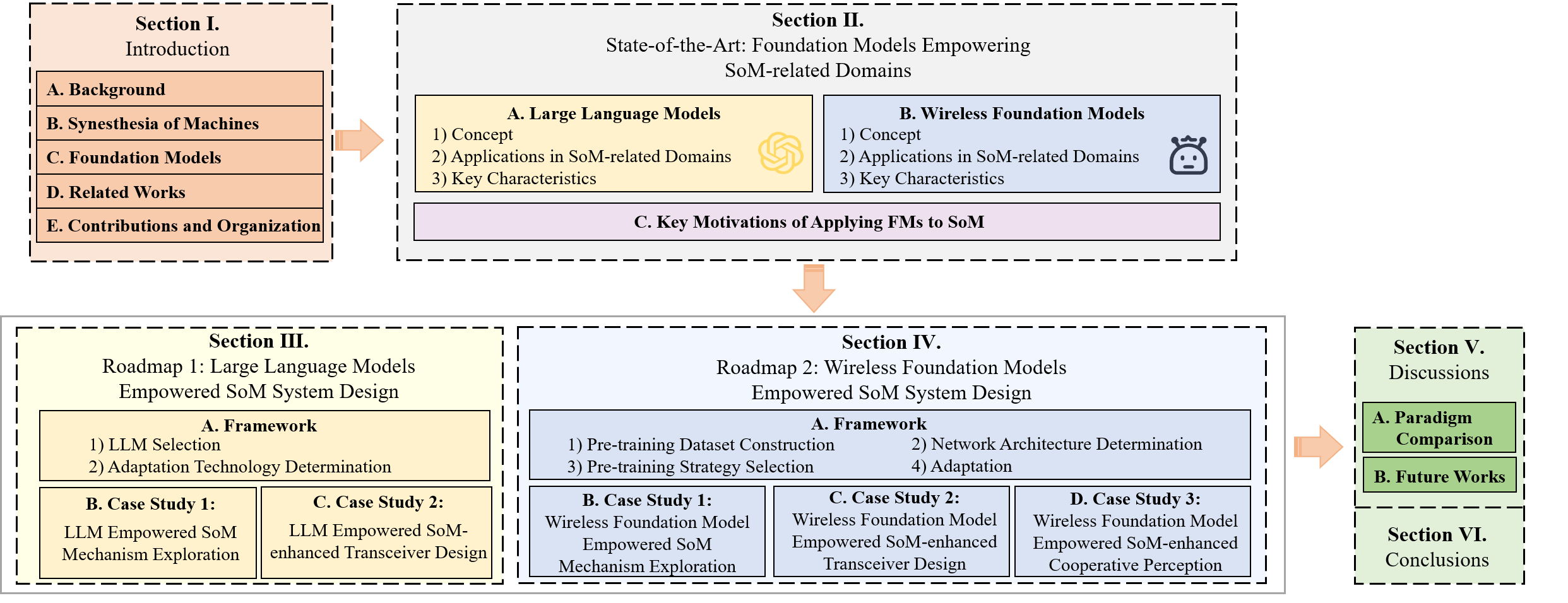}
    \vspace{-3mm}
    \caption{The organization of this paper.}
    \vspace{-5mm}
    \label{organization11}
\end{figure*}

\subsection{Contributions and Organization}
To the best of our knowledge, this paper is the first systematic research on FM-empowered AI-native SoM system design. 
In this paper, inspired by existing studies on SoM-related domains empowered by FMs, we for the first time systematically categorize FMs into two types: general-purpose foundation models, specifically LLMs, and SoM domain-specific foundation models, namely wireless foundation models.
Inspired by the superior capabilities of these two types of FMs in addressing the existing challenges of SoM systems, we propose two novel roadmaps for empowering SoM systems with FMs, i.e., LLM-based and wireless foundation model-based design.
For each roadmap, we first present a detailed step-by-step framework and provide several case studies illustrating its application in SoM system design following the outlined framework.
{For SoM mechanism exploration, we propose LLM-based path loss generation (LLM4PG) and scatterer generation (LLM4SG) scheme, and wireless channel foundation model (WiCo).
For SoM-enhanced transceiver design, we propose LLM-based wireless multi-task SoM transceiver (LLM4WM) and wireless foundation model (WiFo).
For SoM-enhanced cooperative perception, we propose a wireless cooperative perception foundation model (WiPo).
Preliminary simulation results validate the superiority of the proposed FMs empowered schemes.}
Finally, we adequately compare existing AI-empowered SoM system design paradigms and discuss potential future research directions.

As shown in Fig. \ref{organization11}, the paper is organized as follows: Section \uppercase\expandafter{\romannumeral2} introduces the current studies on SoM-related fields empowered by LLMs and wireless foundation models, and further derives several key motivations for applying FMs to SoM. 
Section \uppercase\expandafter{\romannumeral3} systematically illustrates the first research roadmap, which leverages LLMs to empower SoM and introduces some representative case studies, while Section \uppercase\expandafter{\romannumeral4} presents the second roadmap, which utilizes wireless foundation models to empower SoM and also presents some representative case studies. 
Section \uppercase\expandafter{\romannumeral5} comprehensively compares the three AI-empowered SoM system design paradigms and outlines future research directions. Finally, Section \uppercase\expandafter{\romannumeral6} concludes this paper.

\section{State-of-the-Art: Foundation Models Empowering SoM-related Domains}\label{section 2}
In this section, we introduce existing SoM-related research enabled by two types of FMs.
For each type of FMs, we first introduce its fundamental concept, summarize its applications in SoM-related fields, and derive its key characteristics.
Moreover, we summarize the advantages of two FMs in addressing the current challenges of SoM systems.

\subsection{Large Language Models}
\subsubsection{Concept}
LLMs \cite{zhao2023survey} typically refer to language models containing hundreds of billions or more parameters, pre-trained on vast amounts of text data, which possess powerful capabilities in solving natural language tasks.
{In this paper, \textit{LLMs broadly refer to general-purpose FMs, including multi-modal large language models (MLLMs)} \cite{liang2024survey}.}
The underlying network of mainstream LLMs is the transformer \cite{vaswani2017attention} architecture, which excels in long-range dependency modeling, offers strong scalability and is well-suited for hardware parallelization.
Based on the presence of an encoder or decoder, LLM architectures can be categorized into three types \cite{pan2024unifying}: encoder-only, encoder-decoder, and decoder-only.
Following the scaling law \cite{kaplan2020scaling}, LLMs have developed three key emergent abilities: in-context learning, instruction following, and step-by-step reasoning, fundamentally setting them apart from smaller models and signaling the emergence of artificial general intelligence (AGI).

The emergence of LLMs has not only revolutionized natural language processing but also empowered various scientific and engineering fields, including mathematics \cite{ahn2024large}, chemistry \cite{guo2023indeed}, biology \cite{zhang2025scientific}, and software engineering \cite{hou2024large}.
There are two main approaches to applying pre-trained LLMs to domain-specific tasks: fine-tuning \cite{han2024parameter} and prompt engineering \cite{sahoo2024systematic}.
Fine-tuning \cite{han2024parameter} is the process of refining pre-trained LLMs on a specific task or domain by adjusting its parameters to enhance task-specific performance.
According to whether the input and output are in linguistic form, fine-tuning can be classified into linguistic fine-tuning and non-linguistic fine-tuning, with the former also known as instruction tuning \cite{zhang2023instruction}.
Prompt engineering \cite{sahoo2024systematic} is the process of strategically designing and optimizing input prompts to effectively guide LLMs toward generating desired outputs without modifying their underlying parameters. This technique leverages the model’s pre-trained knowledge by structuring prompts in a way that enhances task performance, improves response accuracy, and aligns outputs with specific requirements.
In addition, retrieval-augmented generation (RAG) and tool usage \cite{shen2024llm} are two enhancement techniques for extending the functionality of LLMs.
RAG improves LLMs by retrieving relevant information from external databases, ensuring responses remain accurate and contextually relevant. 
This technique enhances the model’s ability to handle domain-specific queries and keeps its knowledge up to date.
Tool usage allows LLMs to interact with external software, APIs, and computational tools, enabling capabilities beyond text generation. 
By leveraging such tools, LLMs can perform calculations, query structured data, and execute complex tasks efficiently.

\subsubsection{Applications in SoM-related Domains}
We systematically introduce existing works that leverage LLMs to empower SoM-related domains, focusing primarily on SoM dataset construction, SoM-enhanced transceiver design, and SoM-enhanced cooperative perception.

\textit{Multi-modal dataset construction for SoM}:
To address the challenge of limited high-quality training data while preserving user privacy, synthetic data has emerged as a potential solution \cite{goyal2024systematic}. 
Existing research has extensively explored how LLMs facilitate both language-oriented and non-language-oriented data synthesis. For language-oriented data synthesis, LLMs leverage their strong text-generation capabilities to augment datasets such as text classification \cite{li2023synthetic}, dialogue \cite{abdullin2024synthetic}, multilingual commonsense \cite{whitehouse2023llm}, and depression interview transcripts \cite{kang2024synthetic}. Techniques like prompt engineering and chain-of-thought (CoT) guide LLMs to generate domain-specific text, including biomedical named entity recognition \cite{tang2023does} and relation triplet extraction \cite{he2024zero} datasets. In specialized fields such as law \cite{zhou2025lawgpt}, medicine \cite{kumichev2024medsyn}, and mathematical reasoning \cite{fedoseev24constraint}, integrating domain-specific knowledge bases or prior constraints enhances the fidelity of LLM-generated text. 
For non-language-oriented data synthesis, LLMs' semantic understanding enables the generation of data in other modalities, including tables \cite{yang2024enhancing}, images \cite{qin2024diffusiongpt}, videos \cite{cao2024medical}, and time-series \cite{zhou2024geng} data. One approach involves fine-tuning LLMs on specific modalities to learn their distributions and generate synthetic data autoregressively \cite{wang2024harmonic}. Alternatively, LLMs can extract specific semantic features to guide pre-trained generators, such as diffusion models, in generating target-modality data that aligns with textual descriptions \cite{gani2023llm}.
However, existing LLM-empowered synthetic data generation methods have not considered multi-modal sensing and communication dataset generation, making them unsuitable for directly supporting SoM system design.

\textit{SoM-enhanced transceiver design}: 
{Wireless networking is the most relevant research area for SoM-enhanced transceiver design, and its integration with LLMs \cite{zhang2024generative,zhang2024generative1} has been widely explored.}
Based on the type of tasks empowered, existing LLM-driven research for wireless networks \cite{shao2024wirelessllm} can be categorized into language-oriented and non-language-oriented tasks.
Leveraging the powerful language understanding and generation capabilities of LLMs, early studies directly utilize prompt engineering schemes to empower language-oriented telecom tasks, including information synthesis \cite{kotaru2023adapting}, code generation \cite{he2024designing}, transceiver configuration \cite{medaranga2024poster}, and software log analysis \cite{taheri2025domain} in the wireless network domain.
To better facilitate domain knowledge transfer, several studies have attempted to adapt LLMs through instruction tuning for wireless tasks such as telecom question and answer (QnA) \cite{zou2024telecomgpt}, summarizing optimization problems \cite{lin2025empowering}, and network analysis \cite{kan2024mobile}.
To further enhance LLMs' decision-making and task execution capabilities, some instruction tuning empowered studies have integrated RAG \cite{gajjar2025oransight} to leverage domain-specific knowledge in communications and have built LLM agents \cite{xiao2024llm} by invoking APIs to achieve physical layer automation.

On the other hand, several studies leverage LLMs' in-context learning to handle non-linguistic tasks of wireless networks, including power control \cite{zhou2024large1}, symbol detection \cite{abbas2024leveraging}, wireless traffic prediction \cite{hu2024self}, and resource allocation \cite{noh2025adaptive}.
However, since LLMs are not inherently skilled at mathematical reasoning, another mainstream approach is to fine-tune them for better domain adaptation.
The pioneering work LLM4CP \cite{liu2024llm4cp} is the first to fine-tune LLMs in a non-linguistic manner for wireless physical layer tasks, achieving improvements in both channel prediction accuracy and generalization performance.
Based on the same idea, subsequent studies \cite{fan2024csi} have explored the application of LLMs in physical layer tasks such as beam prediction \cite{sheng2025beam}, CSI feedback \cite{cui2025exploring}, channel estimation \cite{xue2025large}.
Moreover, some studies have further explored integrating multi-modal information for networking \cite{wu2024netllm}, while other studies have attempted to fine-tune LLMs for multiple physical layer tasks simultaneously \cite{liu2025llm4wm,zheng2024large}.
Nevertheless, existing LLM-empowered wireless network designs ignore the integration with multi-modal sensing information, making them unsuitable for direct use in SoM transceiver design.

\textit{SoM-enhanced cooperative perception design}:
Efficient and robust information sharing among agents plays a vital role in cooperative perception. 
Deep learning enabled semantic communications have shown great potential to significantly improve transmission efficiency, which only transmits necessary information relevant to the specific task at the receiver \cite{qin2021semantic}. LLMs have been harnessed to enhance semantic communication designs due to their remarkable semantic understanding capabilities, including both language-oriented and non-language-oriented semantic communications. For language-oriented semantic communications, some studies exploit LLMs' summarization and error-correction capabilities to preserve semantic equivalence against wireless channel effects, including language-level \cite{nam2024language} and token-level \cite{guo2023semantic,wang2024large, pokhrel2024large} processing. To further improve the adaptability, some works utilize LLMs to evaluate the importance of features of small models and assign important features to good subchannels, which can be regarded as knowledge distillation from LLMs \cite{jiang2024semantic}. For non-language-oriented semantic communications, LLMs can be integrated with generative models to ensure consistent data generation between transceivers \cite{jiang2024large}. Some studies transmit both compact features of source data and text prompts generated from MLLMs, thereby generating high-fidelity data at the receiver with low data transmission overhead \cite{du2024generative, chen2024semantic, qiao2024latency, zhao2024lamosc}. Besides, LLMs can also address task heterogeneity in multi-user semantic communication systems by employing parameter-efficient fine-tuning methods \cite{chen2024personalizing}. 
However, as LLM-based semantic communication relies on converting source data into linguistic representations, it is not well-suited for universal multi-modal cooperative perception.

\subsubsection{Key Characteristics}
Based on the comprehensive research on LLMs in the SoM-related domain, we can summarize the key characteristics of LLMs in solving SoM-related tasks as follows.
\begin{itemize}
\item \textit{In-context learning}: LLMs exhibit strong in-context learning capabilities, enabling them to learn new tasks from task-specific prompts without fine-tuning. This greatly reduces the need for dedicated fine-tuning datasets and allows quick adaptation from a few examples.

\item \textit{Semantic understanding and generation}: Pre-trained on large-scale textual corpora, LLMs demonstrate deep semantic understanding and high-level abstraction. They can further autoregressively generate coherent, domain-specific content across fields such as literature and code.

\item \textit{General knowledge transfer}: LLMs naturally internalize a vast amount of world knowledge, enabling strong generalization to new tasks and domains with high adaptability and universality.

\end{itemize}
\subsection{Wireless Foundation Models}
\subsubsection{Concept}
In addition to leveraging general-purpose FMs, i.e., LLMs, another approach is to develop domain-specific FMs for the SoM system, termed wireless foundation models in this paper.
{Specifically, \textit{a wireless foundation model is a model pre-trained on broad wireless and multi-modal sensing data (optional), typically at scale using self-supervision, and adaptable to a wide range of SoM-related tasks through few-shot or zero-shot learning}}.
In contrast to task-specific models, the pre-training paradigm and scale effects endow wireless foundation models with distinct capabilities, offering the potential to fundamentally transform SoM system design.
For example, a single model can handle multiple related SoM processing tasks, significantly reducing the number of required models. 
Additionally, when data distribution shifts, wireless foundation model-based approaches enable efficient adaptation with minimal effort, drastically lowering the cost of data collection and model fine-tuning.

{Nevertheless, developing wireless foundation models for SoM systems entails three key challenges: 
\begin{itemize}
\item \textit{Heterogeneity of modalities}: It is challenging to simultaneously process heterogeneous wireless data in SoM systems, such as CSI, channel impulse response (CIR), and IQ (In-phase and Quadrature) signals, along with diverse multi-modal sensing data, including LiDAR point clouds, RGB images, depth maps, and radar point clouds.
\item \textit{Complexity of tasks}: Compared to language tasks, SoM processing involves more complex task types that are difficult to be uniformly modeled as next-token prediction.
\item \textit{Scarcity of datasets}: Unlike readily available language and visual datasets, high-quality SoM datasets \cite{cheng2023m,cheng2025synthsom} are challenging to obtain due to the need for precise alignment between wireless and multi-modal sensing data.
\end{itemize}}

\subsubsection{Applications in SoM-related Domains}
Unlike the extensive use of FMs in CV and NLP, research on wireless foundation models for SoM remains in its early stages. 
We systematically introduce existing studies on domain-specific FMs in SoM-related domains, including SoM dataset construction and SoM-enhanced transceiver design.

\textit{Multi-modal dataset construction for SoM}:
With powerful cross-modal generative abilities, FMs are particularly suited for synthetic data generation. We classify existing work by pre-training strategies, emphasizing methodological distinctions and their impact.
Autoregressive modeling, based on next-token prediction as utilized in UniAudio \cite{yang2023uniaudio}, consolidates various tasks via tokenization, enabling robust task generalization. 
Masked learning methods, such as MaskGIT \cite{chang2022maskgit}, enhance FMs' understanding of image data while enabling parallel decoding for significantly faster image generation. 
Diffusion models \cite{bluethgen2024vision}, leveraging their powerful generative capability and controllability via guidance signals, are widely applied in text-to-image tasks. 
Contrastive learning-based methods, such as SymTime \cite{wang2025mitigating} and VILA-U \cite{wu2024vila}, facilitate efficient cross-modal alignment, improving cross-modal generation. 
Notably, these pre-training paradigms are not strictly distinct but can be integrated for synergistic benefits. 
For instance, MRGen \cite{wu2024mrgen} combines diffusion models with masked modeling for a region-controlled generation.
Nevertheless, the application of FMs for SoM dataset generation remains unexplored.

\textit{SoM-enhanced transceiver design}:
{Due to the powerful representation capabilities, self-supervised learning \cite{yang2025revolutionizing} has been applied to several sensing and communication tasks, including fingerprint localization \cite{salihu2020low}, channel charting \cite{ferrand2021triplet}, wireless power control \cite{naderializadeh2021contrastive}, beam mapping \cite{chafaa2022self}, signal classification \cite{davaslioglu2022self}, spectrum sensing \cite{zhao2025transformer}, channel estimation \cite{zhang2023self}, and geolocation-based MIMO transmission \cite{liu2024leveraging}.}
Nonetheless, these studies remain confined to single-task scenarios. 
Recognizing CSI as a versatile representation for diverse physical-layer tasks underscores the motivation for developing a wireless foundation model for multiple wireless tasks. 
Early work \cite{huangfu2019realistic} first explored a pre-trained realistic channel model for wireless channel data, leveraging a BERT-like architecture and self-supervised pertaining. 
This model's adaptability across pilot contamination mitigation, channel compression, and channel fingerprinting offers preliminary evidence of its generalized understanding of wireless channels.
Furthermore, the Large Wireless Model (LWM) \cite{alikhani2024large} has been proposed to develop a foundation model based on Masked
Channel Modeling (MCM) self-supervised learning for wireless channels and demonstrates its effectiveness across multiple downstream tasks, including cross-frequency beam prediction, LoS/NLoS classification, and robust beamforming.
Nevertheless, LWM can only handle space-frequency two-dimensional CSI and still requires additional fine-tuning for downstream tasks.
Therefore, the first wireless foundation model (WiFo) \cite{liu2024wifo} for channel prediction was proposed to handle 3D CSI, which is pre-trained on extensive diverse CSI datasets and can be directly applied for inference without fine-tuning.
It is the first versatile model capable of simultaneously addressing various channel prediction tasks across diverse CSI configurations and simulation results validate its remarkable zero-shot prediction performance.
Furthermore, \cite{catak2025bert4mimo} explored a BERT-based wireless foundation model for channel prediction, while \cite{guo2025prompt} investigated a prompt-enabled wireless foundation model for CSI feedback.

Several studies have explored building wireless foundation models to jointly handle CSI-related communication and sensing tasks.
A joint-embedding self-supervised method for wireless channel representation learning \cite{salihu2024self} was proposed to learn invariant and compressed channel representations, which is fine-tuned for wireless localization and path loss generation.
A Vision Transformer (ViT)-based radio foundation model \cite{aboulfotouh2024building} employs Masked Spectrogram Modeling (MSM) as a self-supervised learning approach for spectrogram learning and can be fine-tuned for human activity sensing and spectrogram segmentation tasks.
Additionally, several BERT-based multifunctional wireless models \cite{zhao2024finding,zhao2024mining} have been proposed, which are first pre-trained in a self-supervised manner and then fine-tuned for CSI prediction, classification tasks, and WiFi sensing.
A CLIP-based wireless foundation model \cite{jiang2025mimo} was proposed to simultaneously capture the joint feature representation of CSI and CIR and exhibits remarkable adaptability across various CSI-related tasks, including channel identification, positioning, and beam management.
Unlike previous pre-training methods limited to a single CSI data type, a CSI-based multi-modal foundation model \cite{jiao20246g} is introduced, leveraging contrastive learning to align CSI with environmental contexts (BS/UE position and status) and derive task-agnostic CSI representations.
Nevertheless, most existing wireless foundation models for wireless communications and sensing tasks remain limited to a single RF modality and have yet to integrate multi-modal sensing information for SoM-enhanced transmission systems.

\begin{table*}[th]
\centering
\caption{Characteristics of FMs for Addressing the Challenges of Existing SoM System Design}
\label{motivation}
\vspace{-2mm}
\begin{tabular}{c|cc}
\toprule
\multirow{2}{*}{Challenges of existing SoM system design} & \multicolumn{2}{c}{Characteristics}                    \\ \cline{2-3} 
                                                          & \multicolumn{1}{c|}{LLMs} & Wireless foundation models \\ \hline
Scarcity of massive and high-quality datasets & \multicolumn{1}{c|}{\makecell[c]{Powerful semantic understanding and\\ autoregressive data generation capability}} & \multicolumn{1}{c}{Powerful cross-modal generative capability} \\ \hline
Limited modeling capability & \multicolumn{1}{l|}{\makecell[c]{Powerful few-shot modeling ability via\\ in-context learning}} & \multicolumn{1}{l}{Powerful modeling capability following scaling law} \\ \hline
Limited generalization across data & \multicolumn{1}{l|}{\makecell[c]{Strong generalization ability benefit \\ from general knowledge}} & \makecell[c]{One-for-all capability for heterogeneous data\\ and  powerful generalization capability via few-shot\\ and even zero-shot learning} \\ \hline
Limited universality across tasks & \multicolumn{1}{l|}{\makecell[c]{Efficient downstream task transfer \\learning enabled by general knowledge}} & \multicolumn{1}{c}{One-for-all capability for diverse tasks} \\ \bottomrule
\end{tabular}
\vspace{-5mm}
\end{table*}
\subsubsection{Key Characteristics}
\begin{itemize}
\item \textit{One-for-all capability}: Wireless foundation models operate in a one-for-all manner, enabling a single model to handle multiple tasks and heterogeneous data simultaneously, significantly reducing the number of models required for deployment.
\item \textit{Powerful modeling capability}: Following the scaling law, wireless foundation models possess enhanced modeling capabilities to capture complex relationships, enabling them to tackle more challenging SoM processing tasks.
\item \textit{Powerful generalization capability}: Wireless foundation models can generalize across different scenarios and tasks, facilitating few-shot and zero-shot learning, thereby reducing the need for additional retraining overhead.
\item \textit{Powerful generative capability}: Wireless foundation models possess powerful generative capabilities, promising high-quality synthesis of datasets for the integration of communication and multi-modal sensing.
\end{itemize}
\subsection{Key Motivations of Applying FMs to SoM}
In light of the derived key characteristics of the two types of FMs in addressing existing SoM-related tasks, they show great potential in addressing the existing challenges related to SoM system design, as shown in Table \ref{motivation}.
\begin{itemize}
\item For the scarcity of SoM datasets, LLMs are expected to leverage their powerful semantic understanding and autoregressive data generation capabilities to synthesize language and multi-modal data, while wireless foundation models can generate accurately aligned multi-modal sensing-communication datasets through their powerful cross-modal generation ability. 
\item Regarding the challenges in modeling, LLMs possess superior few-shot learning capabilities through in-context learning, while wireless foundation models have a strong ability to handle complex SoM problems by following the scaling law.
\item For data generalization difficulties, LLMs demonstrate strong generalization ability through general knowledge transfer, while wireless foundation models possess powerful multi-dataset joint learning capabilities and impressive zero-shot generalization ability.
\item To address the challenge of limited task universality, LLMs enable effective transfer learning for downstream tasks, whereas wireless foundation models offer enhanced one-for-all capabilities, allowing them to tackle multiple tasks simultaneously.
\end{itemize}
Even so, the pipeline and detailed use cases of applying the two types of FMs to SoM system design are still lacking.
Therefore, in Sections \ref{roadmap1} and \ref{roadmap2}, we propose two roadmaps for FM-empowered SoM system design, utilizing LLMs and wireless foundation models, respectively.
Specifically, the framework and several case studies of each roadmap are illustrated, to provide design guidance for researchers.

\section{Roadmap 1: Large Language Models Empowered SoM System Design}\label{roadmap1}
\begin{figure*}[th]
    \centering
    \includegraphics[width=0.95\linewidth]{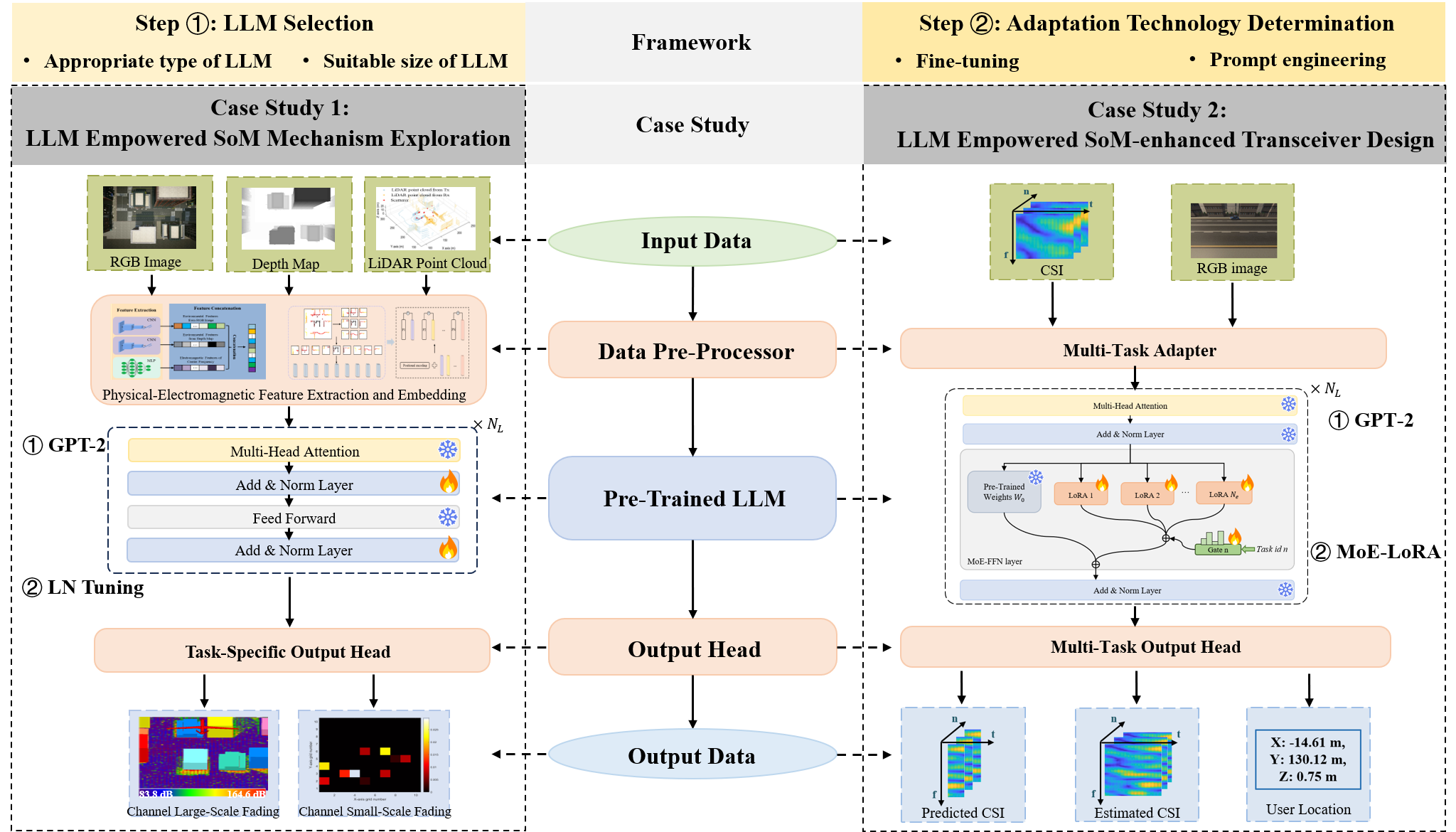}
    \caption{An illustration of the framework of roadmap 1 and the proposed schemes for the two case studies introduced.}
    \vspace{-3mm}
    \label{roadmap1-fig}
\end{figure*}

In this section, we present the first roadmap for designing a foundation model-empowered SoM system, which harnesses the capabilities of pre-trained LLMs via fine-tuning or prompt engineering.
We begin by introducing a framework that highlights two core components of SoM system design: LLM selection and adaptation technology determination. 
We then present two case studies to demonstrate its practical implementation.
The overall framework and the detailed designs of the two case studies are shown in Fig. \ref{roadmap1-fig}.

\subsection{Framework}
\subsubsection{LLM Selection}
The first step in tackling a specific SoM problem with LLMs is to determine the choice of LLM backbone, as it plays a foundational and decisive role in the task's overall performance.
Two key aspects need to be carefully considered, i.e., the type and the size of the LLM.

\begin{itemize}
\item \textit{Appropriate type of LLM}: 
To better leverage the LLM's understanding capabilities and facilitate general knowledge transfer, the type of LLM should closely align with the nature of the SoM task.
In SoM tasks characterized by temporal dependencies, such as multi-modal sensing-assisted channel prediction, FMs designed for time-series data are more appropriate than language-based LLMs.
For instance, study \cite{xue2025large} reformulates the delay-Doppler (DD) domain channel prediction problem as a time series prediction task of DD-domain parameters, demonstrating that time-series FMs like Timer \cite{liu2024timer} and Time-Mixture of experts (MoE) \cite{shi2024time} can be directly applied and subsequent fine-tuning on specific vehicular channel data further improves prediction accuracy.

\item \textit{Suitable size of LLM}: 
The model size \cite{zhang2024scaling} is another critical factor that directly impacts both the performance and the deployment feasibility of SoM systems.
Larger LLMs offer stronger generalization and cross-modal reasoning abilities, making them ideal for complex SoM tasks involving high-dimensional, multi-modal data. However, their high computational and memory demands limit real-time applicability in resource-constrained environments.
Conversely, smaller LLMs \cite{liu2024llm4cp} provide lower latency and are easier to fine-tune for domain-specific tasks, making them more suitable for real-time inference. 
{In addition, it is noted that the performance of this scheme initially improves and then gradually levels off as the size of the LLM increases.
For instance, for networking optimization based on LLM fine-tuning \cite{wu2024netllm}, an LLM with 1 billion parameters is already sufficient.}
Techniques like knowledge distillation \cite{xu2024survey} and quantization can further reduce large model sizes while preserving their core capabilities, striking a balance between efficiency and performance.

\end{itemize}
\subsubsection{Adaptation Technology Determination}
Once the appropriate LLM is selected, the next step is to determine the adaptation technology for the specific SoM task. 
The choice of adaptation method significantly influences the system's effectiveness and generalization. 
Broadly, the adaptation strategies can be categorized into two approaches: fine-tuning and prompt engineering.
\begin{itemize}
\item \textit{Fine-tuning}:
Fine-tuning updates the LLM's internal parameters on SoM-specific data, enhancing its cross-domain transferability for specific SoM tasks.
It includes both non-linguistic fine-tuning, which adapts the LLM to process non-textual SoM data, and instruction tuning, which enhances its ability to follow SoM domain-specific instructions.
Although fine-tuning requires additional labeled data, it achieves superior performance over task-specific models in few-shot scenarios.
For example, in cross-frequency generalization tests, the LLM-based channel prediction scheme \cite{liu2024llm4cp} requires only 30 CSI samples to outperform model-based schemes \cite{yin2020addressing}, whereas other task-specific models need more than 100 CSI samples.
\item \textit{Prompt engineering}:
Unlike fine-tuning, prompt engineering guides the LLMs to perform SoM tasks through carefully designed prompts, avoiding the need for additional parameter updates.
By preserving the original parameters of the LLMs, prompt engineering fully leverages the model's inherent semantic understanding capabilities for multi-modal information fusion in SoM systems.
For instance, in SoM feature transmission tasks, multi-modal LLMs can map diverse data types (e.g., text, images, and LiDAR point clouds) into a unified semantic space \cite{jiang2024large}, enabling more efficient encoding and transmission.
To overcome the LLM's limitations in numerical reasoning and cross-domain adaptation, RAG and external tool integration enable access to external databases and APIs for enhanced accuracy and effectiveness.
These capabilities make prompt engineering a lightweight yet powerful adaptation technique for SoM systems, particularly when large-scale fine-tuning is impractical.

\end{itemize}

\subsection{Case Study 1: LLM Empowered SoM Mechanism Exploration}
As the theoretical foundation of SoM research, it is essential to explore the complex and nonlinear SoM mechanism, i.e., the mapping relationship between communications and sensing \cite{huang2024lidar,huang2024scatterer}.
On one hand, we explore the SoM mechanism between sensing and path loss, and further propose LLM4PG as an LLM-based scheme for path loss generation.
On the other hand, we explore the SoM mechanism between sensing and multipath fading, and further propose LLM4SG as an LLM-based scheme for scatterer generation \cite{han2025llm4sp}. The framework of the proposed LLM4PG/LLM4SG scheme, i.e., case study 1, is given in Fig.~\ref{roadmap1-fig}.

\begin{itemize}
\item \textit{Step 1: LLM Selection.} In the SoM mechanism exploration, we need to select an appropriate LLM as the backbone from two perspectives, including task adaptability and model performance/efficiency. For task adaptability, existing pre-trained LLMs are not specifically designed for SoM mechanism exploration tasks. In this case, we select a mature LLM with sufficient generality and extensibility. For the model performance/efficiency, the model generalization capability in transfer learning and its inference overhead are considered to meet the high-precision requirement of SoM mechanism exploration tasks. Therefore, we select the lightweight GPT-2 \cite{radford2019language} as the backbone in the LLM4PG and LLM4SG schemes.
\item \textit{Step 2: Adaptation Technology Determination.}  Since the SoM mechanism exploration is non-language-oriented, the fine-tuning approach is more suitable for addressing these challenges. On one hand, we bridge the significant gap between the natural language domain and the multi-modal information domain. For RGB-D images, we extract physical environment features and employ feature-level fusion to map these features into the natural language domain. For LiDAR point clouds, we perform voxelization preprocessing followed by patch partitioning and positional encoding to convert the data into tokens compatible with the LLM feature space. Meanwhile, an output module transforms the GPT-2 generated token sequences into required channel fading information, including path loss in the LLM4PG scheme and  scatterers in the LLM4SG scheme.
On the other hand, we adopt the fine-tuning strategy where most pre-trained LLM parameters remain frozen, and further transfer general knowledge to the SoM mechanism exploration task. {Specifically, we employ LN Tuning \cite{qi2022parameter}, where only the LayerNorm parameters are set as trainable to reduce computational overhead while preserving model generality.}
\end{itemize}

\begin{figure}[t]
    \centering
    \includegraphics[width=0.99\linewidth]{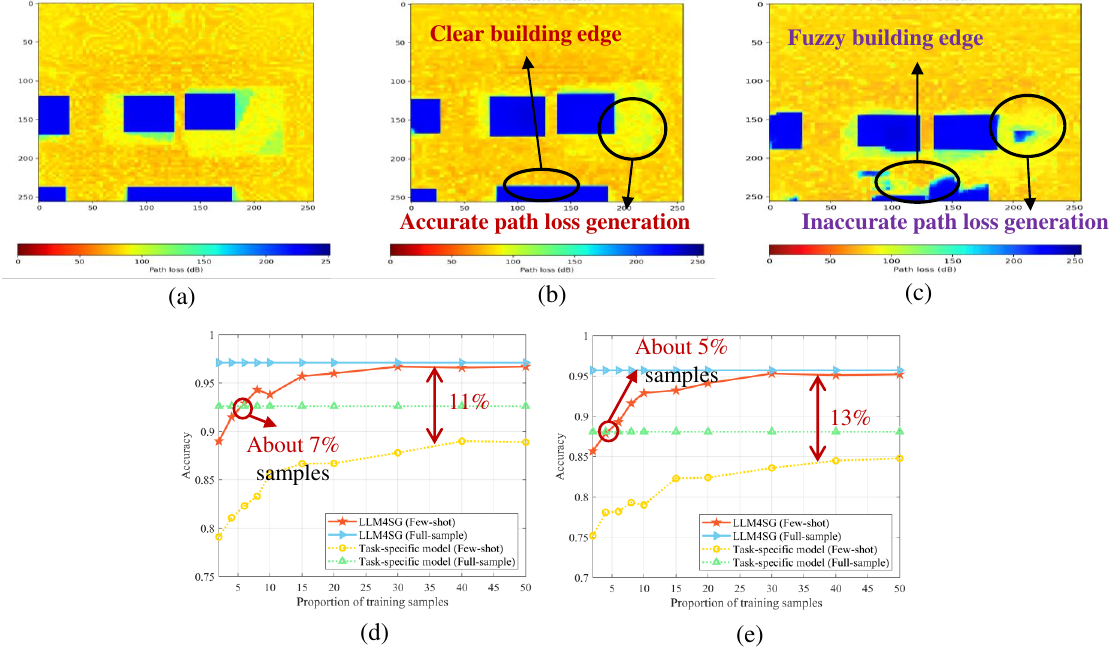}
    \vspace{-2mm}
    \caption{Performance of SoM mechanism exploration. (a) Ray-tracing-based path loss map. (b) Path loss map result via the LLM4PG scheme. (c) Path loss map result via the GAN-based scheme. (d) Scatterer number results via the LLM4SG scheme and the ResNet-based scheme. (e) Scatterer location results via the LLM4SG scheme and the ResNet-based scheme.}
    \vspace{-5mm}
    \label{LLM_mapping}
\end{figure}

The proposed LLM4PG and LLM4SG schemes are trained on the SynthSoM dataset \cite{cheng2023m}.
The advantage of the proposed LLM4PG and LLM4SG schemes is shown in Fig.~\ref{LLM_mapping}. For the proposed LLM4PG scheme, Figs.~\ref{LLM_mapping}(a)--(c) show that the LLM, i.e., GPT-2, demonstrates superior accuracy over the task-specific model, i.e., generative adversarial network (GAN), in the exploration of SoM mechanism between sensing and path loss. Compared with the GAN-based scheme,  the proposed LLM4PG scheme demonstrates higher accuracy in reconstructing building edges in path loss maps, particularly for fading caused by buildings. For the proposed LLM4SG scheme, Figs.~\ref{LLM_mapping}(d) and (e) evaluate the generalization capability of the LLM, i.e., GPT-2, and the task-specific model, i.e., ResNet, in the exploration of the SoM mechanism between sensing and small-scale fading. 
The model is first trained on sub-6 GHz samples and then fine-tuned with a subset of 28 GHz samples for transfer testing. The proposed LLM4SG scheme, through knowledge transfer, achieves over 11\% higher generalization accuracy compared to the ResNet-based scheme, reaching the performance of the ResNet-based scheme trained on the full dataset with about 7\% of the training samples. 
{For all case studies in this paper, the inference time is evaluated on the same machine with an NVIDIA GeForce RTX 4090 GPU.}
{Then, the complexity of the aforementioned scheme, i.e., the number of parameters and the average inference time, is given in Table~\ref{llm-cost11}. It can be seen from Table~\ref{llm-cost11} that the LLM4PG and LLM4SG schemes show an inference speed closely matching that of task-specific models, i.e., GAN-based and ResNet-based schemes.} 

\begin{table}[]
\centering
\caption{The Number of Network Parameters (Training Parameters/Total Parameters) and the Interference Time per Batch (Batch Size is Set to 8) of Case Study 1 for Roadmap 1}
\label{llm-cost11}
\vspace{-2mm}
\begin{tabular}{c|c|c}
\toprule
& Parameters (M) & Inference time (ms) \\ \hline
LLM4PG &    {52.23/275.70}            & 9.90                   \\   \hline
GAN &    {45.61/45.61}            & 7.36                   \\   \hline
LLM4SG &    {5.08/86.19}           & 7.96                   \\   \hline
ResNet &    {23.53/23.53}           & 5.67                   \\    
\bottomrule
\end{tabular}
\vspace{-5mm}
\end{table}

 \begin{table}[]
\centering
\caption{The Number of Network Parameters (Training Parameters/Total Parameters) and the Interference Time per Batch (Batch Size is Set to 8) of Case Study 2 for Roadmap 1}
\label{llm-cost21}
\vspace{-2mm}
\begin{tabular}{c|c|c}
\toprule
& Parameters (M) & Inference time (ms) \\ \hline
SM-STL &    {1.92/1.92}           & 1.16                   \\   \hline
SM-MTL &    {1.92/1.92}            & 1.81                   \\   \hline
LLM4WM &    {2.20/84.10}           & 8.73                   \\   \hline
SM-STL-RGB &    {2.29/2.29}           & 3.17                   \\  \hline  
SM-MTL-RGB &    {2.29/2.29}           & 3.30                   \\   \hline
LLM4WM-RGB &   {4.50/98.10}           & 9.08                   \\    
\bottomrule
\end{tabular}
\vspace{-5mm}
\end{table}

\subsection{Case Study 2: LLM Empowered SoM-enhanced Transceiver Design}
We consider several common vision‐aided tasks in SoM transceiver design, including channel estimation \cite{soltani2019deep}, prediction \cite{liu2024llm4cp}, and user positioning \cite{salihu2022attention}. A multi-task learning approach is employed to exploit the synergy among these SoM tasks, thereby maximizing the spectral efficiency of the communication system. We will elaborate on how to effectively design an LLM-based wireless multi-task SoM transceiver by following the two fundamental steps outlined above.
\begin{itemize}
\item \textit{Step 1: LLM Selection}. We first need to select a suitable LLM as the backbone based on task requirements, considering both the type and size of the LLM. As existing pre-trained LLMs are not specifically tailored to the characteristics of SoM tasks, we adopt a relatively mature LLM. Furthermore, while large-scale LLMs generally excel in transfer learning and zero-shot generalization, their high dimensionality and deep architectures introduce inference overhead that is unacceptable for physical layer tasks. Therefore, we choose the lightweight GPT-2 \cite{radford2019language} as the backbone.
\item \textit{Step 2: Adaptation Technology Determination}. Since the selected tasks are non-language-oriented, the fine-tuning approach is more suitable for addressing these challenges. Specifically, the Mixture of experts with low-rank adaptation (MoE-LoRA) \cite{liu2025llm4wm} method is employed for multi-task fine-tuning, effectively mitigating task conflicts while enabling efficient transfer of GPT-2’s general knowledge. {To align the LLM's output with both the label dimensions and the feature space, task-specific adapters with a ResNet-style \cite{he2016deep} architecture are introduced at the output stage.}
\end{itemize}
\begin{figure}[t]
    \centering
    \includegraphics[width=0.8\linewidth]{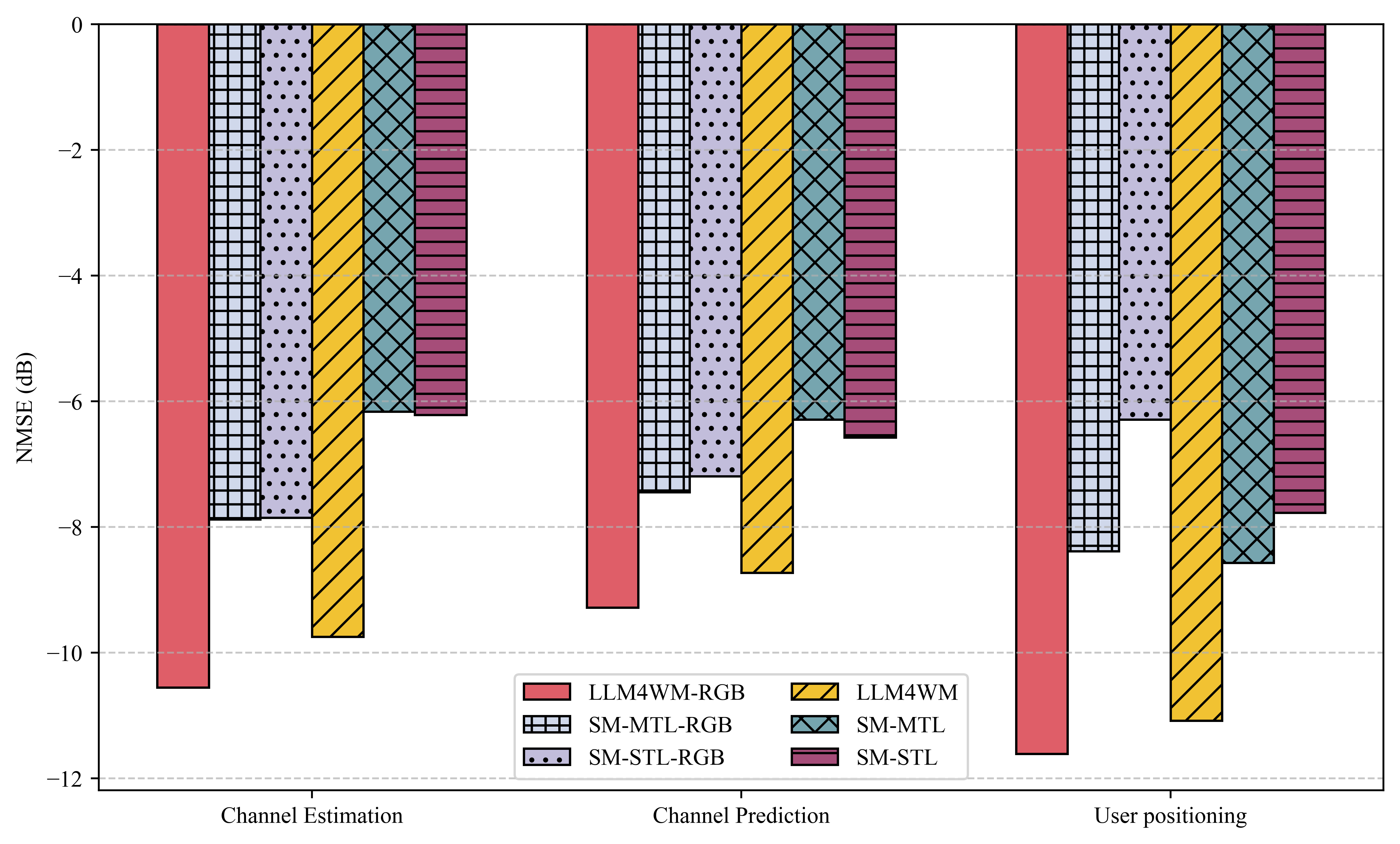}
    \vspace{-2mm} 
    \caption{\textcolor{black}{NMSE performance comparison of the proposed LLM4WM scheme with the SM-MTL and SM-STL schemes across three SoM-related tasks.}}
    \vspace{-5mm}
    \label{result-case-study-1}
\end{figure}

Based on the above two steps of analysis, the framework of LLM4WM is illustrated in Fig.~\ref{roadmap1-fig}.
Then, we employ the SynthSoM \cite{cheng2025synthsom} dataset to acquire aligned RGB images, vehicle Global Positioning System (GPS) data, and CSI.
 
The advantages of the proposed LLM4WM scheme are illustrated in Fig. \ref{result-case-study-1}, where comparisons are made between the specialized model single-task learning (SM-STL) scheme and the specialized model multi-task learning (SM-MTL) scheme. 
Furthermore, to compare the gains from vision assistance, we conducted experiments for each scheme both with and without vision input, and the vision-assisted variants are denoted with the suffix “-RGB”, while the versions without vision input retain the original scheme name.
The Cross-Stitch network \cite{misra2016cross} is adopted for the SM-MTL scheme, while the SM-STL scheme also follows the configuration in \cite{misra2016cross}, maintaining the same network architecture as the multi-task learning scheme but being trained and tested on individual tasks.
It can be observed that vision-assisted schemes exhibit consistent performance gains relative to their non-vision-assisted counterparts, underscoring the contribution of visual information to transmission tasks. Furthermore, constrained by limited modeling capacity, the SM-MTL-RGB scheme is unable to leverage joint multi-task information, resulting in performance comparable to that of SM-STL-RGB. By contrast, the LLM4WM-RGB framework achieves state-of-the-art (SOTA) performance across all tasks, attributed to the extensive general knowledge encoded within LLM, thereby exhibiting superior task generalization.
We also evaluate the complexity of each scheme in terms of the number of parameters and the average inference time across the three tasks, as summarized in Table~\ref{llm-cost21}. Notably, LLM4WM-RGB demonstrates an inference speed closely matching that of specialized models.
Although only three SoM-related tasks are selected in this case study, the LLM-based scheme is capable of handling a larger number of tasks, achieving even better multi-task learning performance, and enhancing the efficiency of utilizing the general knowledge embedded in LLMs \cite{liu2025llm4wm}.

\section{Roadmap 2: Wireless Foundation Models Empowered SoM System Design}\label{roadmap2}
\begin{figure*}[th]
    \centering
    \includegraphics[width=1\linewidth]{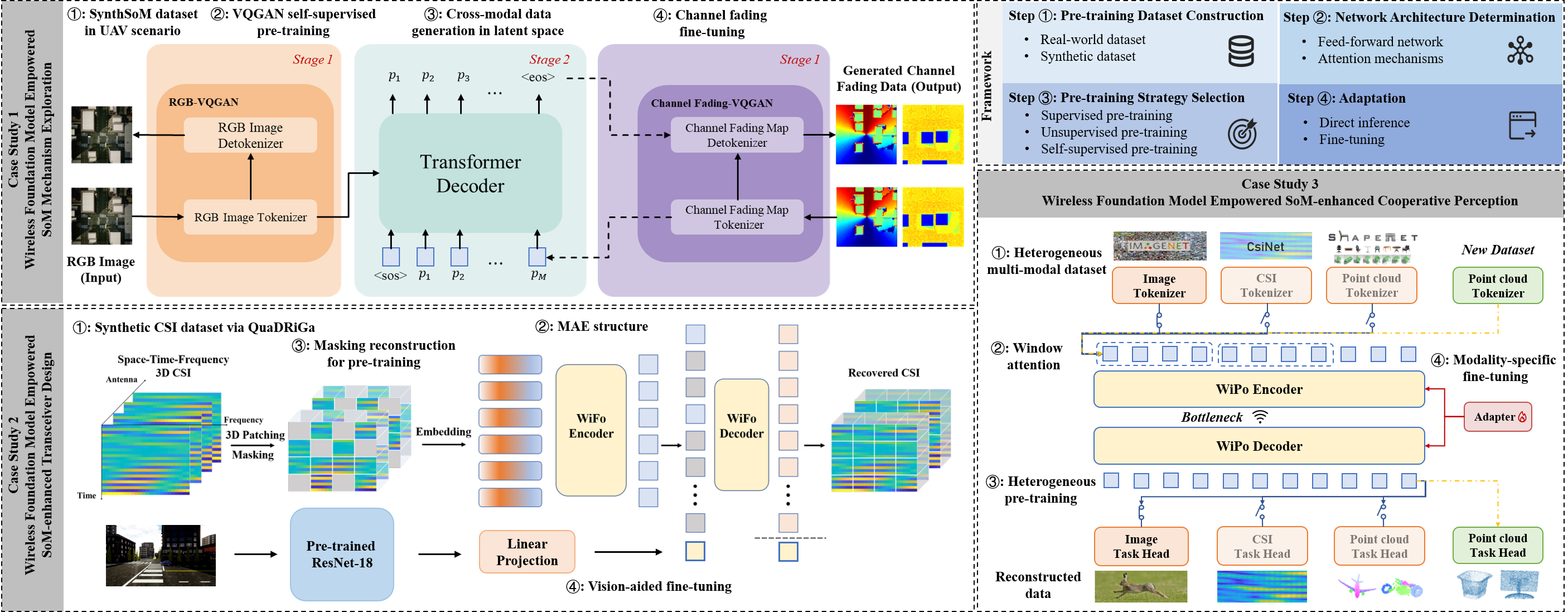}
    \vspace{-2mm}
    \caption{An illustration of the framework of roadmap 2 and the proposed schemes for the three case studies introduced.}
    \vspace{-5mm}
    \label{roadmap2-fig}
\end{figure*}

In this section, we introduce the second roadmap, which is to build a wireless foundation model from scratch tailored to specific SoM domains.
We present a framework with four key steps: dataset construction, network architecture, pre-training strategy, and adaptation, followed by three case studies on building wireless foundation models for specific SoM tasks.
The overall framework of the second roadmap and the network processing of the three case studies are shown in Fig. \ref{roadmap2-fig}.

\subsection{Framework}
\subsubsection{Pre-training Dataset Construction}
The scale and quality of the dataset determine the performance upper bound of the wireless foundation model built from scratch for the SoM system \cite{liu2024datasets}.
Unlike readily available language and visual datasets, the construction of datasets for SoM systems needs to collect precisely aligned communication and multi-modal sensing data from the same physical environment, such as RF communication data, e.g., channel matrices and path loss, RF sensing data, e.g., mmWave radar point clouds, and non-RF sensing data, e.g., RGB images, depth maps, and LiDAR point clouds, resulting in huge challenge. 
{In general, the existing multi-modal sensing-communication dataset for SoM can be classified into the real-world dataset obtained via measurement equipment, the synthetic dataset collected via simulation software, and the generated dataset via artificial intelligence generated content (AIGC) models based on  the SoM mechanism.}
\begin{itemize}
\item \textit{Real-world dataset:} With the help of measurement equipment, the real-world dataset is of high accuracy. A measurement
dataset, named KITTI, for multi-modal sensing was developed in \cite{geiger2013vision}. The calibrated and synchronized RGB image, depth map, and LiDAR point cloud were collected in the KITTI dataset. To further include  communication channel data, a measurement dataset, named
DeepSense 6G, was constructed in \cite{alkhateeb2023deepsense}, including multi-modal sensing data, e.g., RGB image and LiDAR point cloud, and communication channel data under the sub-6 GHz band and mmWave band. Although the real-world dataset facilitates the algorithm validation, it is of huge difficulty to customize multi-modal sensing and communication scenarios attributed to the cost concern.
\item \textit{Synthetic dataset:} Attributed to the limitations of the real-world dataset, many synthetic datasets have been constructed as a supplement to the real-world dataset. With the help of accurate software, multi-modal sensing and communication data can be efficiently collected in the synthetic dataset. The synthetic dataset ViWi was constructed in \cite{alrabeiah2020viwi}, including channel information, RGB images, depth maps, and LiDAR point clouds. However, mmWave radar data was ignored in \cite{alrabeiah2020viwi}. To overcome this limitation, a synthetic dataset, named  M$^3$SC, was developed in \cite{cheng2023m}, which contained multi-modal sensing and communication data in various vehicular scenarios. To further include unmanned aerial vehicle (UAV) scenarios, a real-world data injected synthetic multi-modal sensing-communication dataset, named SynthSoM, was developed in \cite{cheng2025synthsom}. The main limitation of the synthetic dataset is accuracy constraints due to some unrealistic assumptions in the collection software.
\item \textit{Generated dataset:} Another approach to construct the multi-modal sensing-communication dataset for SoM is based on AIGC models in conjunction with SoM mechanisms. For the generation of multi-modal sensory data, extensive effective AIGC models can be utilized, which have the ability to leverage generative architectures to autonomously generate high-quality and diverse multi-modal sensory data by learning underlying data distributions \cite{aiello2022cross,zheng2023autofed}. For the generation of communication channel data, the explored SoM mechanism can be utilized to achieve efficient and high-fidelity cross-modal generation of
communication channel data, which is temporally and spatially consistent with the multi-modal sensory data \cite{han2025llm4sp}. 
The primary limitation in the generated dataset lies in the dependency of multi-modal data generation accuracy on both the AIGC model performance and the precision of SoM mechanism exploration.

\end{itemize}
\subsubsection{Network Architecture Determination}
Transformer \cite{vaswani2017attention} has revolutionized deep learning by enabling models to capture long-range dependencies with high computational efficiency. Its self-attention mechanism allows for parallel processing of sequences, making it particularly effective for pre-training on large-scale data. Therefore, Transformer has become the mainstream architecture for LLMs, underpinning advancements in natural language processing \cite{radford2019language}, computer vision \cite{dosovitskiy2020image}, robotics \cite{wang2024scaling} and related fields. However, SoM tasks differ from classical tasks and require network architectures to be adapted to address their unique characteristics. Potential areas for improvement primarily include the feed-forward network (FFN) and attention mechanisms.

\begin{itemize}
\item \textit{Feed-forward network}: To further improve modeling capacity and representation capability, some studies replace the FFN with other networks. A typical approach involves using a Mixture of Experts (MoE) to substitute the FFN, thereby achieving superior multi-task performance with faster inference speed, such as switch transformer \cite{fedus2022switch}, mixtral of experts \cite{jiang2024mixtral} and deepseek-MoE \cite{dai2024deepseekmoe}. Another representative variant is Gated Linear Units (GLU) \cite{shazeer2020glu} applied in the LLaMa \cite{touvron2023llama}, which is straightforward to implement and yields superior performance. By employing structures with enhanced representational capabilities, such as MoE, the model can better learn the interrelationships among various SoM tasks, thereby achieving improved generalization and performance.

\item \textit{Attention mechanisms}: The attention module is a key block in the transformer, enabling the model to capture complex contextual relationships between input tokens. However, it also accounts for a significant portion of the computation load in the transformer. Many works propose new attention mechanisms to improve computational efficiency, including flash attention \cite{dao2022flashattention}, grouped-query attention \cite{ainslie2023gqa}, window attention \cite{liu2021swin} and so on. Beyond efficiency, appropriate attention mechanisms can also enhance model performance. For example, the Swin Transformer introduces stronger inductive biases tailored to visual tasks \cite{liu2021swin}. In the context of SoM tasks, which require both efficient communication and accurate sensing, timeliness is often critical. Therefore, optimizing the attention mechanism represents a promising direction for balancing task performance with runtime efficiency.

\end{itemize}
\subsubsection{Pre-training Strategy Selection}
Selecting an appropriate pre-training strategy is crucial for model performance, as different pre-training approaches directly affect the model's ability to learn from datasets, thereby influencing its performance across various SoM tasks. Based on the type of supervision, pre-training strategies can be categorized into three types: supervised pre-training, unsupervised pre-training, and self-supervised pre-training.

\begin{itemize}
\item \textit{Supervised pre-training}:
Supervised pre-training learns directly from labeled, high-quality datasets, enabling the model to acquire strong task-specific capabilities. For example, ResNet \cite{he2016deep} is pre-trained on the large-scale image dataset ImageNet \cite{deng2009imagenet}, which contains over 1 million labeled images across 1,000 categories, enabling the model to learn rich hierarchical visual representations and extract high-level semantic features. However, supervised pre-training heavily depends on massive labeled datasets, which incurs high annotation costs, and the performance of the pre-trained model is also affected by the quality of the labels.

\item \textit{Unsupervised pre-training}:
Unsupervised pre-training enables the model to learn meaningful representations of data by leveraging rule-based proxy tasks, without requiring manually annotated labels. It is commonly used for dimensionality reduction, clustering, and has been applied to various SoM-related tasks, such as multi-user precoding \cite{guo2024deep}. Specifically, the multi-user precoding network trained with unsupervised learning does not require a ground truth precoding matrix. Instead, it directly optimizes spectral efficiency by learning the intrinsic representations of the precoding task, thereby improving its generalization capability.

\item \textit{Self-supervised pre-training}:
Self-supervised pre-training generates training labels by leveraging the inherent structure or information within the data. Examples include next token prediction \cite{radford2019language}, masked modeling \cite{he2022masked}, and contrastive learning \cite{chen2020simple}. This approach also eliminates the need for manual annotations while providing excellent scalability and zero-shot generalization capabilities. For instance, In \cite{jiao20246g}, contrastive learning is utilized to align CSIs with associated environment descriptions characterized by BS positions, enabling the well-aligned representations to be directly applied in downstream classification tasks, such as LoS/NLoS identification.
\end{itemize}

\subsubsection{Adaptation}
To ensure the wireless foundation model effectively generalizes across diverse SoM tasks and scenario conditions, adaptation techniques must be carefully designed. The adaptation strategies primarily include direct inference and fine-tuning, which determine how the model transfers knowledge to unseen tasks and scenarios.

\begin{itemize}
\item \textit{Direct inference}:
{Compared with LLM-based solutions, the wireless foundation model exhibits strong zero-shot generalization ability, meaning that the wireless foundation model could generalize to new data distributions without any additional labeled data.} 
This is particularly useful for dynamically changing environments where collecting task-specific training data is impractical. 
By leveraging pre-trained representations, wireless foundation models for SoM can infer relevant patterns from unseen data distributions. 
For instance, \cite{liu2024wifo} demonstrated that large-scale pre-trained models exhibit strong zero-shot capabilities for channel prediction.

\item \textit{Fine-tuning}:
Fine-tuning further enhances the model’s adaptability by allowing it to learn new data distributions or even new tasks quickly with minimal labeled examples. This is particularly beneficial for SoM scenarios where labeled data is scarce or expensive to acquire. Additionally, parameter-efficient tuning (PEFT) techniques like LoRA \cite{wu2024netllm} and MoE-LoRA \cite{liu2025llm4wm} facilitate efficient knowledge transfer while minimizing computational overhead, making them well-suited for real-time adaptation in SoM systems.
\end{itemize}

{The deployment and adaptation of wireless foundation models for downstream tasks can be further enhanced via a cloud-edge-terminal collaborative architecture \cite{zhang2024cloud}. 
In this framework, large-scale wireless foundation models are centrally maintained and periodically updated in the cloud, leveraging global knowledge and computational resources. 
To enable efficient on-device inference and adaptation, techniques such as knowledge distillation and parameter-efficient fine-tuning are employed to compress and transfer the core capabilities of cloud models to lightweight edge or terminal models. 
Moreover, federated learning can facilitate privacy-preserving model adaptation by allowing terminals to locally update model parameters based on their data, with only the aggregated model updates being sent to the cloud. 
This collaborative paradigm ensures continuous model evolution, low-latency adaptation, and efficient resource utilization, thereby supporting scalable and secure deployment of wireless foundation models in heterogeneous real-world environments.}

\subsection{Case Study 1: Wireless Foundation Model Empowered SoM Mechanism Exploration}
\begin{table}[]
\centering
\caption{The Number of Network Parameters (Training Parameters/Total Parameters) and the Interference Time per Batch (Batch Size is Set to 8) of Case Study 1 for Roadmap 2}
\label{wifo-cost22}
\vspace{-2mm}
\begin{tabular}{c|c|c}
\toprule
& Parameters (M) & Inference time (ms) \\ \hline
WiCo-PG &    29.21/70.56            & 5.92                   \\   \hline
LLM-based &    16.36/63.86            & 5.26                   \\   \hline
GAN &    45.61/45.61            & 4.36                   \\   \hline
WiCo-MG &    25.87/76.85           & 7.24                   \\   \hline
LLM-based &    12.39/68.48            & 6.51                   \\   \hline
ResNet &   43.09/43.09             & 5.29 
\\  \bottomrule
\end{tabular}
\vspace{-5mm}
\end{table}
To explore the complex and nonlinear SoM mechanism between sensing and communications, we propose the wireless channel foundation model for the first time. {Given the scarcity of channel data compared to sensory data, by exploring the SoM mechanism, the WiCo scheme leverages more accessible sensory data to achieve efficient and high-fidelity cross-modal generation of channel data.} The proposed WiCo scheme contains two parts, including WiCo for path loss generation (WiCo-PG)  and WiCo for multipath component generation (WiCo-MG). For clarity, we elaborate on the proposed WiCo-PG scheme, with the WiCo-MG scheme adopting a similar methodology. 
The framework of the proposed WiCo-PG and WiCo-MG schemes, i.e., case study 1, is shown in Fig.~\ref{roadmap2-fig}.

\textit{Step 1: Pre-training Dataset Construction}.
Based on the requirement of the WiCo-PG scheme, we investigate the existing multi-modal sensing-communication dataset with diverse scenarios. Considering the volume and quality of the dataset, we utilize the SynthSoM dataset in \cite{cheng2025synthsom} to develop the WiCo-PG scheme. The SynthSoM dataset covers various air-ground multi-link cooperative scenarios with diverse data modalities, such as the RF communication, i.e., 140K sets of 
channel matrices and 18K sets of path loss, RF sensing, i.e., 136K sets of mmWave radar waveforms with 38K 
radar point clouds, and non-RF sensing, i.e., 145K RGB images, 290K depth maps, as well as 79K sets of LiDAR point clouds.

\textit{Step 2: Network Architecture Determination}.
To select a proper network architecture of  the WiCo-PG scheme, the accuracy and generalization need to be considered. To explore the SoM mechanism between RGB images and path loss maps, we enhance  the Pathways Autoregressive Text-to-image (Parti) generative model architecture proposed by Google, which integrates the visual discrete representation of the VQGAN network with the autoregressive generation capability of the transformer.


\textit{Step 3: Pre-Training Strategy Selection}.
To select a proper pre-training strategy, it is essential to adopt self-supervised learning tailored for the accurate exploration of SoM mechanism between sensing and path loss. Through data augmentation and noise suppression training methods, we can effectively extract intrinsic data features and enhance the robustness of the proposed WiCo-PG scheme. 

\textit{Step 4: Adaptation}.
For the model adaptation, the focus lies in achieving a smooth transition of the pre-trained WiCo-PG to new datasets by training a small number of parameters. Furthermore, the pre-trained WiCo-PG can be fine-tuned on different scenarios and frequency bands by efficiently leveraging the pre-trained knowledge of channel fading generation tasks.  Then, model parameters are optimized for the accurate exploration task of SoM mechanism between sensing and path loss, thus enhancing domain-specific accuracy.

\begin{figure}[t]
    \centering
    \includegraphics[width=0.6\linewidth]{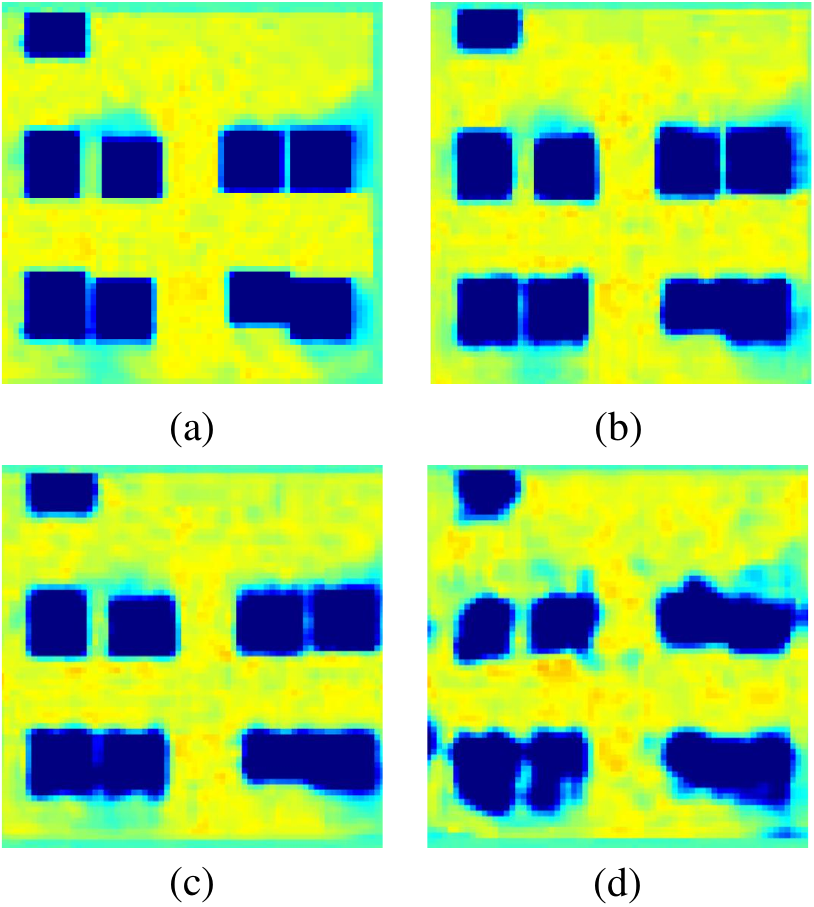}
    \vspace{-2mm}
    \caption{Comparisons of path loss map results. (a) Ray-tracing-based result. (b) The WiCo-PG scheme. (c)  The LLM-based scheme. (d) The GAN-based scheme.}
    \vspace{-4mm}
    \label{result-pl}
\end{figure}

The WiCo-PG scheme is trained on the SynthSoM dataset \cite{cheng2025synthsom} with path loss maps under 28 GHz frequency band and  UAV images. 
Figs.~\ref{result-pl}(a) and (b) demonstrate a close agreement between the ray-tracing-based path loss map and the generated path loss map via the WiCo-PG scheme through the powerful cross-modal generation ability of FMs. Specifically, the generated path loss map via the WiCo-PG scheme accurately identifies and reconstructs building contours, and further precisely generates path loss caused by buildings. Compared with the LLM-based scheme based on GPT-2 in Fig.~\ref{result-pl}(c), the WiCo-PG scheme achieves over 3\% higher accuracy in path loss generation. Furthermore, a huge difference between the ray-tracing-based path loss map and the generated path loss map via the task-specific model, i.e., GAN, can be observed in Figs.~\ref{result-pl}(a) and (d), where path loss generation is blurred on building contours. The parameter size and inference time of the aforementioned schemes are listed in Table~\ref{wifo-cost22}. The WiCo-PG scheme and the GAN-based scheme have the same order of magnitude regarding storage and computational overhead.

{Similar to the WiCo-PG scheme, the WiCo-MG scheme also contains four main steps, including pre-training dataset construction, network architecture determination, pre-training strategy selection, and adaption. For the pre-training dataset construction, our constructed SynthSoM dataset \cite{cheng2025synthsom} in the UAV scenario is utilized. For the network architecture determination, we utilize VQGAN and Transformer attributed to its visual discretized representation and autoregressive generation capability. For the pre-training strategy selection, we also exploit the self-supervised learning tailored for the high-fidelity and efficient multipath parameter generation. For the model adaption, the pre-trained network is fine-tuned on different scenarios and frequency bands by leveraging the pre-trained knowledge efficiently.  As shown in Table~\ref{wifo-cost22}, the WiCo-MG scheme and the ResNet-based scheme are also with the same order of magnitude regarding storage and computational overhead.

\begin{figure}[t]
    \centering
    \includegraphics[width=0.6\linewidth]{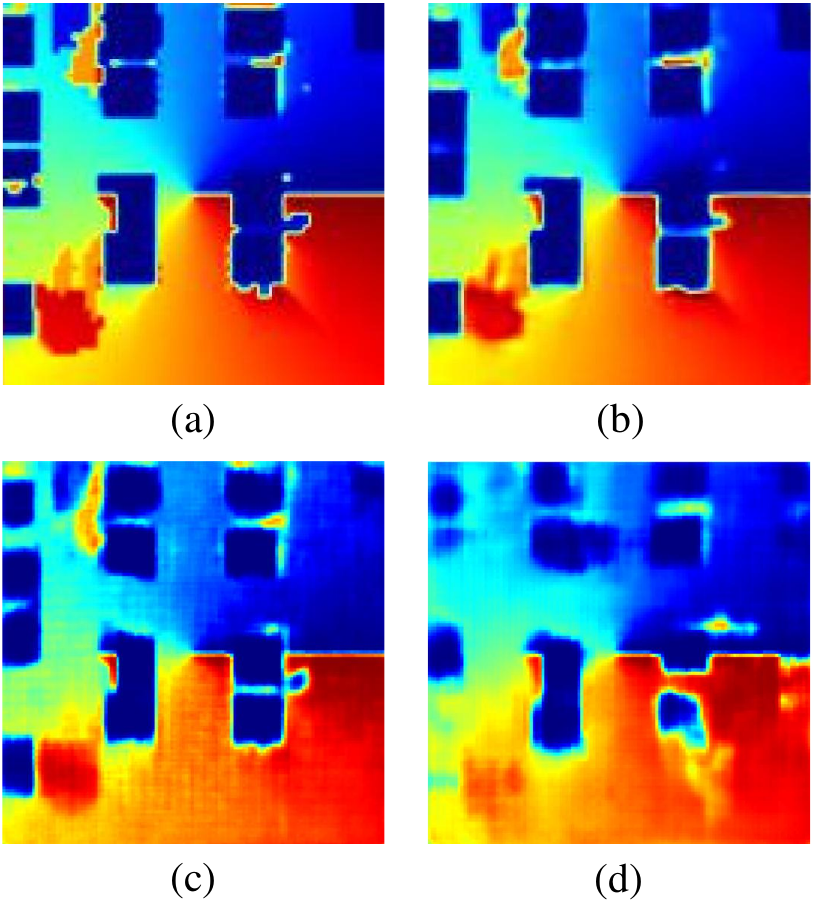}
    \vspace{-2mm}
    \caption{Comparisons of multipath parameter results. (a) Ray-tracing-based result. (b) The WiCo-MG scheme. (c) The LLM-based scheme. (d) The ResNet-based scheme.}
    \vspace{-4mm}
    \label{result-mp}
\end{figure}

The WiCo-MG scheme is trained on the SynthSoM dataset \cite{cheng2025synthsom} with multipath parameters, i.e., departure of angle (DoA), under 28 GHz frequency band and  UAV images. Figs.~\ref{result-mp}(a) and (b) show a close consistency between the ray-tracing-based DoA and the generated DoA via the WiCo-MG scheme through the efficient cross-modal generation ability of FMs. Compared with the LLM-based scheme based on GPT-2 in Fig.~\ref{result-mp}(c), the WiCo-MG scheme in Fig.~\ref{result-mp}(b) achieves over 5\% higher accuracy in multipath parameter generation. In addition, in Figs.~\ref{result-mp}(a) and (d), the ray-tracing-based DoA and the generated DoA via the task-specific model, i.e., ResNet, are exceedingly different, where the building edges are blurred and the DoA in the occluded regions behind buildings is notably inaccurate.

}

\subsection{Case Study 2: Wireless Foundation Model Empowered SoM-enhanced Transceiver Design}\label{case-study-2.1}
\begin{table}[]
\centering
\caption{The Number of Network Parameters (Training Parameters/Total Parameters) and the Interference Time per Batch (Batch Size is Set to 8) of Case Study 2 for Roadmap 2}
\label{wifo-cost}
\vspace{-3mm}
\begin{tabular}{c|c|c}
\toprule
                 & Parameters (M) & Inference time (ms) \\ \hline
WiFo (zero-shot) &   0/21.60             & 8.74                    \\ \hline
WiFo (CSI)       &  2.16/21.60              & 8.74                    \\ \hline
WiFo (CSI+RGB)   &  2.68/33.71              &    10.90                 \\ \hline
LLM4WM (CSI)       &  2.20/84.10              &   8.67                   \\ \hline
LLM4WM (CSI+RGB)   &  4.50/98.10              &    9.08                 \\ \hline
Task-specific (CSI)     &  14.84/21.60              &    8.74                 \\ \hline
Task-specific (CSI+RGB) &  15.36/33.71              & 10.90                    \\ \bottomrule
\end{tabular}
\vspace{-5mm}
\end{table}
In this case study, we consider vision-aided frequency-domain channel prediction powered by wireless foundation models. 
Due to the limited scale of existing multi-modal sensing and communication datasets, it is challenging to pre-train a multi-modal wireless foundation model from scratch. 
Therefore, we first build a CSI-oriented wireless foundation model, termed WiFo \cite{liu2024wifo}, and then fine-tune it using vision data for frequency-domain channel prediction in new scenarios, as shown in Fig. \ref{roadmap2-fig}.

\begin{itemize}
\item \textit{Step 1: Pre-training Dataset Construction.}
CSI datasets can be obtained through real-world measurements, ray-tracing simulations, and statistical channel modeling. 
Existing measurement datasets \cite{yaman2024luvira,dichasus2021} for channel prediction are limited in scale and diversity, constraining the performance of pre-trained models. 
While ray-tracing simulations offer flexibility, they come with high computational costs. 
Therefore, we leverage the QuaDRiGa channel generator to generate a large-scale 3D CSI dataset compliant with 3GPP standards, containing over 160k samples. The dataset covers 16 heterogeneous scenarios and system configurations, with more details provided in \cite{liu2024wifo}.

\item \textit{Step 2: Network Architecture Determination.}
It is worth noting that CSI data and video are quite similar in type, both being structured and continuous 3D data.
Inspired by the success of masked autoencoders (MAE) in image and video pre-training, we propose an MAE-based network for CSI reconstruction.
As shown in Fig. \ref{roadmap2-fig}, diverse CSI data is first transformed into varying token numbers via 3D patching and embedding, facilitating processing by transformer blocks. 
For both the encoder and decoder, we introduce a novel positional encoding structure (STF-PE) to capture the 3D positional information.

\item \textit{Step 3: Pre-training Strategy Selection}.
Noticing that the CSI prediction and the masking reconstruction pre-training task are similar, we adopt a masking-based self-supervised pre-training approach to enable WiFo with general reconstruction capabilities. 
Specifically, in addition to random masked reconstruction, we also design time and frequency domain masked reconstruction pre-training tasks to enhance the model's ability for both the time and frequency domain channel prediction.
\item \textit{Step 4: Adaptation}.
The pre-trained WiFo can be directly used or fine-tuned for frequency-domain channel prediction on specific scenarios.
We consider a fine-tuning approach with visual information enhancement, utilizing aligned RGB and CSI sample pairs.
{Since WiFo is designed to handle 3D CSI, we first concatenate the 2D CSI along the time dimension to transform it into a three-dimensional format.}
Specifically, the visual information, processed by the pre-trained ResNet-18 \cite{he2016deep}, is mapped to a token through a fully connected layer and concatenated to the input tokens of the WiFo decoder.
During fine-tuning, only the additional fully connected layer of ResNet-18, the first layer of the WiFo decoder, and the final output layer are trainable, while the rest of the network is frozen to retain general knowledge.
\end{itemize}
\begin{figure}[t]
    \centering
    \includegraphics[width=0.75\linewidth]{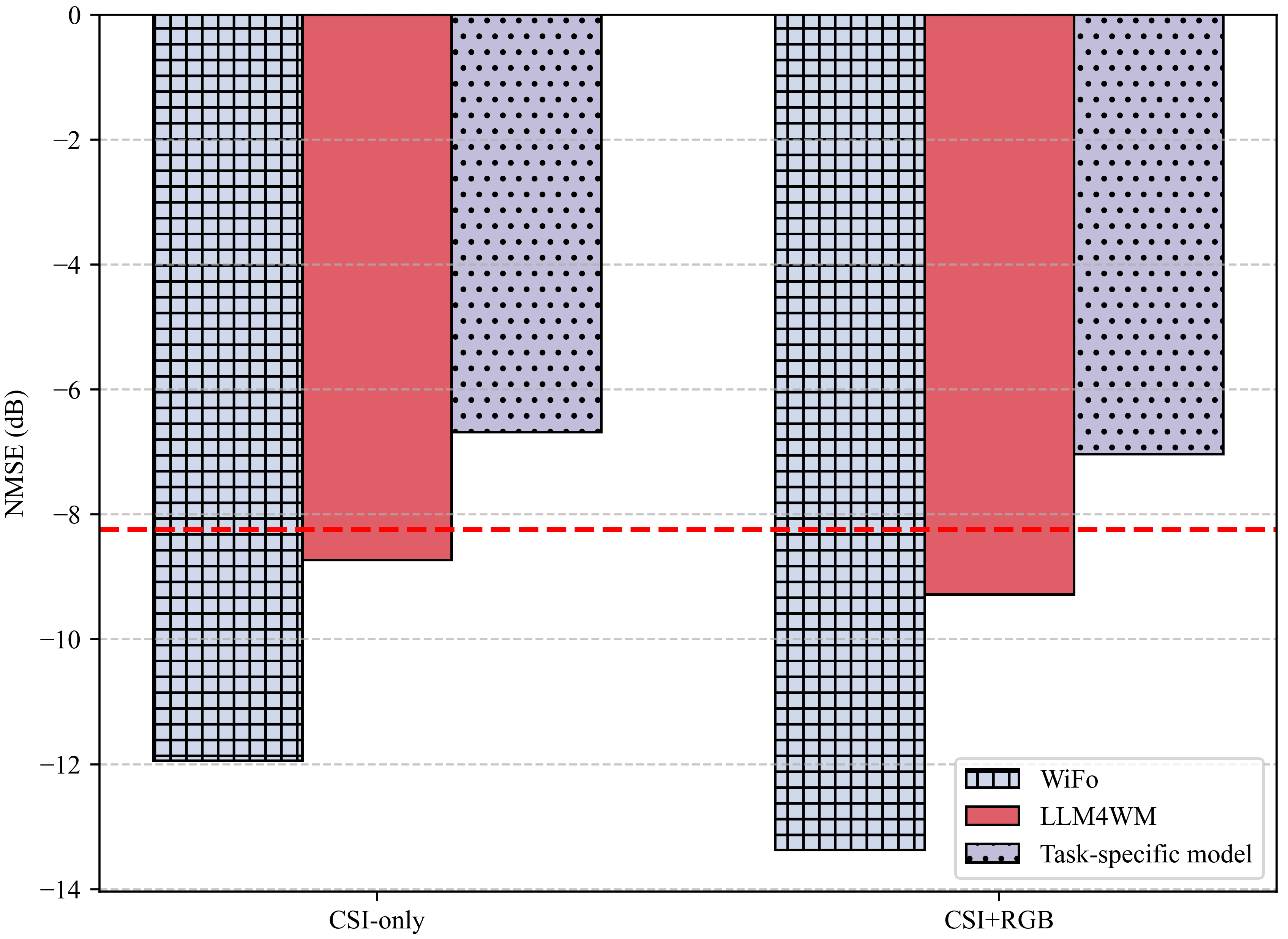}
    \vspace{-2mm}
    \caption{NMSE performance comparison of WiFo-based, LLM-based, and task-specific model-based schemes.}
    \vspace{-4mm}
    \label{result-wifo}
\end{figure}

{We utilize the SynthSoM dataset \cite{cheng2025synthsom} for fine-tuning, which includes 1,500 aligned CSI and RGB sample pairs with 16 antennas and 64 subcarriers.} 
We aim to predict the CSI of the successive 32 subcarriers based on the first 32 subcarriers, using NMSE as the metric for prediction accuracy. 
For the task-specific models used as baselines, the WiFo encoder is randomly initialized and participates in training, while the other components are the same as those in the WiFo-based scheme.
Simulation results of WiFo-based, LLM-based, and task-specific model-based schemes are illustrated in Fig. \ref{result-wifo}.
It can be observed that whether using only CSI or combining it with RGB, the WiFo-based approach significantly outperforms the task-specific approach and the LLM-based scheme. 
This indicates that the pre-trained wireless foundation model has stronger few-shot learning capabilities than task-specific models and can quickly adapt to specific scenarios and demonstrates its superior performance compared to the LLM-based scheme.
Furthermore, the zero-shot performance of WiFo achieves better results than the task-specific models and performance comparable to LLM4WM, suggesting that even without fine-tuning, WiFo can still deliver acceptable performance in new scenarios. 
The parameter size and inference time of the above schemes are shown in Table \ref{wifo-cost}. 
It can be observed that the increase in model parameters and inference time brought by the introduction of visual information is relatively limited, demonstrating the feasibility of applying multi-modal sensing information to practical transceiver design.

\subsection{Case Study 3: Wireless Foundation Model Empowered SoM-enhanced Cooperative Perception}

Given the massive and diverse data transmission demands in multi-agent communication networks, it is essential to develop FMs capable of supporting multi-modal data transmission for cooperative perception. 
Compared to training separate models for each modality, FMs offer greater generality and scalability, substantially reducing the cost and complexity of deployment.
To this end, we propose a wireless cooperative perception foundation model, named WiPo, which supports modality-agnostic feature transmission, as illustrated in Fig. \ref{roadmap2-fig}.
\begin{itemize}
\item \textit{Step 1: Pre-training Dataset Construction.}
Modality-agnostic feature transmission aims to learn shared encoding rules across heterogeneous data. 
To achieve this, data diversity is essential during pre-training. By leveraging existing open-source datasets from various modalities, we construct a heterogeneous multi-modal dataset. 
In each pre-training iteration, a sample is randomly selected from this dataset to promote generalization across modalities.

\item \textit{Step 2: Network Architecture Determination}.
For modality-agnostic feature transmission, network architecture selection must consider both universality and performance. Vanilla transformers, such as ViT, offer strong universality due to their flexibility with respect to input token lengths and dimensionalities. 
We further adopt a window-based attention mechanism to capture detailed features, as used in the Swin Transformer.

\item \textit{Step 3: Pre-training Strategy Selection}.
The foundation model consists of three components: lightweight tokenizers, a unified backbone, and specific task heads. The modality-specific tokenizers and task heads introduce inductive biases tailored to each modality and generate tokens compatible with the shared backbone. During pre-training, different tokenizers and task heads are employed for each modality, while the backbone remains shared across all modalities. The primary objective of pre-training is to obtain a unified backbone capable of reconstructing heterogeneous multi-modal data under the influence of wireless channel distortions. A similar heterogeneous pre-training strategy has been applied in the field of robotic manipulation \cite{wang2024scaling}.

\item \textit{Step 4: Adaptation}.
Once a unified backbone is pre-trained, it can be directly applied to new datasets and even unseen modalities for feature transmission. Adaptation requires training only a small number of parameters in modality-specific tokenizers and task heads. The backbone can either be frozen or fine-tuned using PEFT techniques. Specifically, we insert adapter modules \cite{houlsby2019parameter} into the transformer layers to leverage the pre-trained knowledge effectively, as shown in Fig. \ref{roadmap2-fig}.
\end{itemize}

\begin{figure}[t]
    \centering
    \vspace{-1mm}
    \subfloat[]
    {\includegraphics[height=0.6\linewidth]{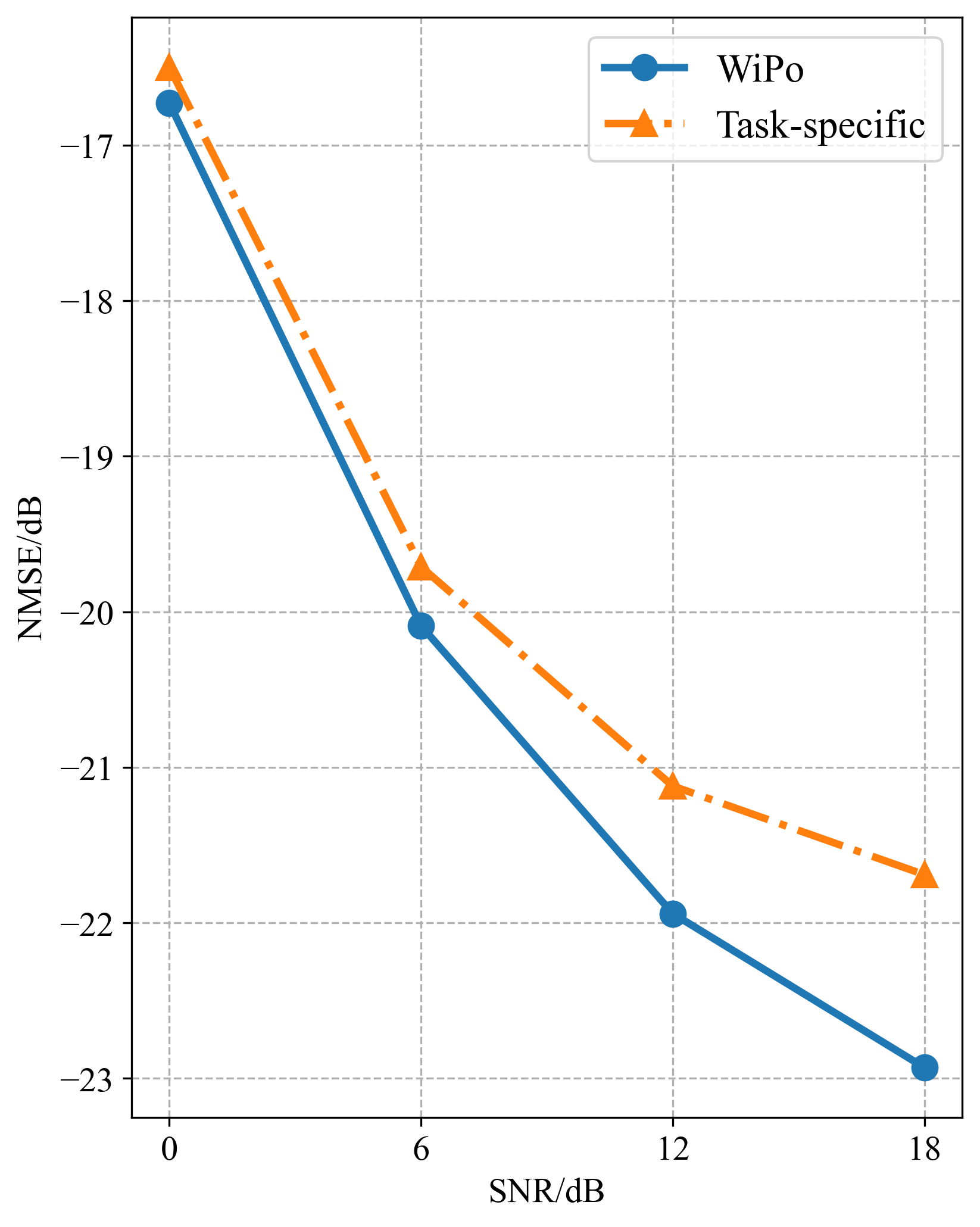}} 
    \subfloat[]{\includegraphics[height=0.6\linewidth]{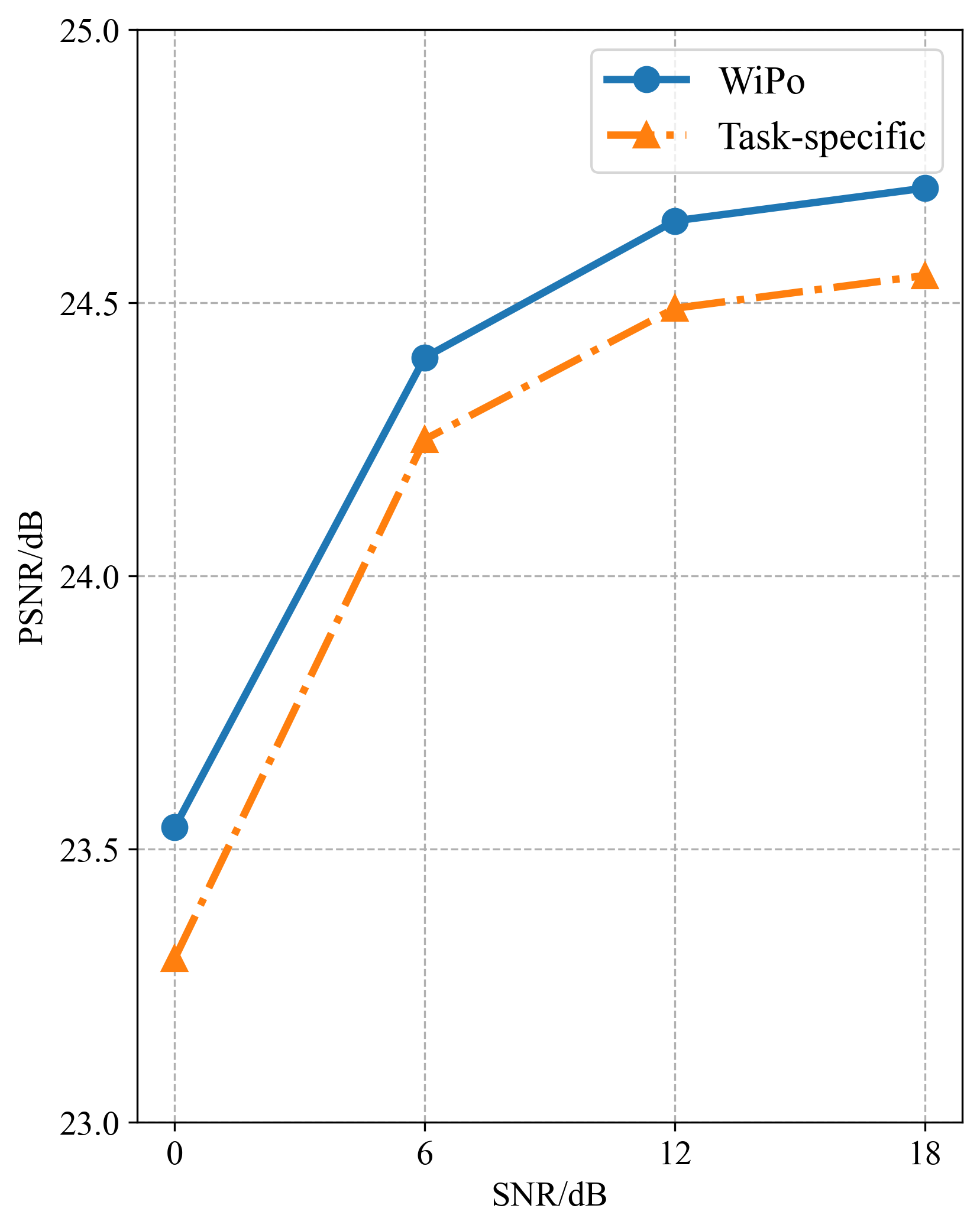}}
    \vspace{-1mm}
    \caption{Reconstruction performance comparison of WiPo and task-specific models. (a) NMSE performance of CSI feedback. (b) Peak signal-to-noise ratio (PSNR) performance of image transmission.}
    \vspace{-3mm}
    \label{result-som-feature}
\end{figure}

{We pre-train the model on large-scale heterogeneous multi-modal datasets, including ImageNet, CsiNet-Outdoor and ShapeNet, and then fine-tune it on the SynthSoM dataset \cite{cheng2025synthsom}.
For comparison, we adopt the same network architecture for baseline models but train them individually for the respective modality. Simulation results for CSI feedback and image transmission tasks are presented in Fig. \ref{result-som-feature}. Across all SNR levels, WiPo consistently outperforms the task-specific models, demonstrating its superior one-for-all and generalization capabilities.}
For inference efficiency, WiPo does not incur additional inference costs, as it shares the same network architecture as task-specific models.
{As shown in Table \ref{WiPo-cost}, the number of stored parameters in WiPo is only 13.65M across three tasks, which is 59.1\% less than that of task-specific models. Moreover, adaptation for new datasets only needs to train a small number of parameters and freeze the pre-trained backbone, demonstrating the flexibility and versatility of WiPo.
WiPo achieves competitive performance with fewer parameters by sharing backbone weights, demonstrating its effectiveness in multi-modal data transmission scenarios.}

\begin{table}[]
\centering
\caption{The number of network parameters (Training Parameters/Total Parameters) of Case Study 3 for Roadmap 2}
\label{WiPo-cost}
\vspace{-2mm}
\begin{tabular}{c|c|c}
\toprule
Parameters (M)              &   WiPo  &  Task-specific \\ \hline
Image Transmission          &   1.03/11.69    &    11.18/11.18       \\ \hline
CSI Feedback                &   0.92/11.58    &    10.98/10.98       \\ \hline
Point Cloud Transmission    &   1.04/11.70    &    11.19/11.19       \\ \hline
\textbf{Stored Parameters}           &   \textbf{13.65}    &    \textbf{33.35}       \\ \bottomrule
\end{tabular}
\vspace{-3mm}
\end{table}

\section{Discussions}
In this section, we first summarize and compare the three existing AI-empowered SoM system design paradigms, including task-specific AI models and the two proposed FM-empowered schemes.
In addition, several open issues and potential directions for future research on FM-empowered SoM system design are discussed.
\begin{table}[t]
\renewcommand{\arraystretch}{1.5}
\caption{Comparison of Three AI-empowered SoM System Design Paradigms, i.e., Task-specific AI Models, LLMs, and Wireless Foundation Models}
\label{ai_comparison}
\centering
\begin{tabular}{c|c|c|c}
\toprule
 & \textbf{\makecell{Task-specific \\AI models}} & \textbf{LLMs} & \makecell{\textbf{Wireless} \\ \textbf{foundation model}} \\
\hline
\makecell[c]{Parameters} & Small & Large & Medium \\
\hline
\makecell[c]{Inference time} & Low & Medium or low & Medium or low \\
\hline
\makecell[c]{Pre-training \\ requirement} & No & No & Yes \\
\hline
\makecell[c]{Modeling \\ capability} & Weak & Medium & Strong \\
\hline
\makecell[c]{Generalization} & Weak & Medium & Strong \\
\hline
\makecell[c]{Universality} & Weak & Medium & Strong \\
\hline
\makecell[c]{Generative \\capability} & Weak & Medium & Strong \\
\bottomrule
\end{tabular}
\vspace{-5mm}
\end{table}
\subsection{Paradigm Comparison of AI-empowered SoM System Design}
As shown in Table \ref{ai_comparison}, we comprehensively compare the three AI-empowered SoM system design paradigms from several perspectives.
It can be observed that the proposed two foundation model-based schemes have significant advantages in terms of modeling capability, generalization, universality, and generative capability. 
Moreover, wireless foundation models not only outperform LLMs in the above aspects but also have fewer parameters, which helps reduce storage overhead.
Although wireless foundation models require an additional training process, the training is offline and does not incur extra overhead during the actual deployment process.
In addition, existing simulations \cite{liu2024llm4cp,liu2024wifo} show that compared to task-specific models, LLMs-based and wireless foundation models-based schemes do not significantly increase inference latency, making them promising for application in real-time systems.

In practical SoM systems, the choice of scheme should align with system requirements and hardware capabilities. 
Task-specific models are ideal for SoM problems with lower task difficulty, such as multi-modal sensing-aided beam prediction \cite{zhang2024integrated} in low-speed scenarios, and are well-suited for deployment on user-side devices with limited computational power. 
Pre-trained LLM-based schemes are better suited for SoM problems of moderate complexity, like frequency-domain channel prediction \cite{liu2024llm4cp} in high-speed scenarios, and are more suitable for deployment on base stations with ample storage and computing resources. 
Wireless foundation model-based schemes excel in addressing high-difficulty SoM challenges, such as zero-shot channel prediction \cite{liu2024wifo}, while requiring fewer computational resources than LLM-based models.
\subsection{Future work}
In this part, future work is categorized by research direction, highlighting key considerations for SoM system design empowered by foundation models.
\subsubsection{Multi-modal dataset construction for SoM} Since the dataset scale and quality determine the ultimate performance limit of AI-native systems, it is necessary to construct a massive and high-quality multi-modal sensing-communication dataset for SoM research. Towards this objective,  based on the generation ability of foundation models, massive multi-modal sensing-communication data can be generated efficiently. To further ensure the quality of generated multi-modal sensing-communication data, it is essential to conduct real-world data injection via digital twin to efficiently guide the process of data generation. As a consequence, models trained on the generated data can be directly deployed in the real world, thus achieving zero-shot generalization.

\subsubsection{SoM mechanism exploration} For the LLM-empowered SoM mechanism exploration, although fine-tuning is an intuitive approach for SoM mechanism exploration, prompt engineering techniques are yet to be utilized to explore SoM mechanisms.
Leveraging such methods, based on the LLM, the SoM mechanism exploration can be explored more accurately. For the WiCo empowered scheme, by collecting real-world data, SoM mechanisms between communications and multi-modal sensing can be accurately explored through real-world injection in diverse scenarios, various frequency bands, and different conditions.

\subsubsection{SoM-enhanced transceiver design}
On one hand, for LLM-empowered SoM transceiver design, a key direction is developing transceivers that retain LLM language capabilities through prompt engineering or other innovative methods. 
{This approach significantly enhances the utilization of LLMs, enabling them to function both as AI service providers and as tools for optimizing transceiver design and improving communication quality, thereby significantly increasing the feasibility of practical deployment.} 
On the other hand, for the wireless foundation model empowered scheme, one promising approach is to integrate multi-modal sensing information, as demonstrated in case study 1 of Section \ref{case-study-2.1}. 
However, since multi-modal sensing information is not considered during the pre-training phase, the wireless foundation models pre-trained solely on CSI data lack native capabilities for aligning and jointly processing multi-modal information, restricting their performance on SoM tasks.
In future research, it is necessary to consider constructing multi-modal native wireless foundation models for SoM transceiver design, which explores the general mechanisms of multi-modal sensing-assisted transceiver design and demonstrates great generalization and universality.
\subsubsection{SoM-enhanced cooperative perception}
On one hand, for LLM-empowered SoM-enhanced cooperative perception, although many pretrained LLMs exist for perception, they largely ignore communication constraints. As the number of collaborating agents increases, limited bandwidth becomes the bottleneck for perception performance. Therefore, it is significant to design communication-efficient LLMs for cooperative perception, which can achieve both communication bandwidth efficiency and reliable perception capability. 
On the other hand, for wireless foundation models supporting data transmission in collaborative perception, existing approaches primarily focus on coordinating heterogeneous tasks, typically under the assumption of single-input single-output (SISO) communication over AWGN or Rayleigh channels. In reality, communication systems feature diverse physical-layer configurations, such as varying numbers of subcarriers, antennas, and users. Therefore, it is imperative to develop novel, physical-layer-aware pretraining strategies that can adapt to these heterogeneous configurations.
\subsubsection{Network system support for SoM} LLMs can serve as intelligent orchestrators that interpret dynamic network states and provide data-driven, real-time decisions for complex resource allocation tasks. 
Beyond language models, specialized foundation models could learn from large-scale network data to intelligently schedule tasks and dynamically reconfigure network parameters across diverse sensing-computing-communication workloads. 
For example, graph-oriented or physics-informed generative models (e.g., diffusion networks) could be developed as dedicated base models that capture network topologies and physical constraints, enabling near-optimal control strategies through sampling from learned solution distributions. 
Integrating such generative models into decision pipelines opens new possibilities for automated service placement, adaptive topology management, and cross-layer optimization, ultimately yielding a self-optimizing, AI-driven network system that fully realizes the SoM paradigm.

\section{Conclusions}
In this paper, we conducted a comprehensive study of FM-empowered SoM system design and established a complete theoretical framework.
In light of existing FM-empowered SoM-related studies, we categorize FMs for SoM system design into general-purpose FMs, i.e., LLMs, and domain-specific FMs, i.e., wireless foundation models. 
In light of this, we identified the key motivations for leveraging FMs to address the existing challenges in SoM systems and for the first time proposed corresponding two research roadmaps, i.e., LLMs-based and wireless foundation model-based SoM system design.
For the first roadmap, we introduced the proposed design framework empowered by LLMs, including LLM selection and adaptation technology determination, and then presented two case studies.
{Specifically, we proposed LLM4PG and LLM4SG for SoM mechanism exploration and LLM4WM for SoM-enhanced transceiver design.}
Similarly, for the second roadmap, we gave a framework to illustrate the specific steps involved in building a wireless foundation model from scratch, including pre-training dataset construction, network architecture determination, pre-training strategy selection, and adaptation. 
{Furthermore, we presented WiCo, WiFo, and WiPo, for SoM mechanism exploration, SoM-enhanced transceiver design, and SoM-enhanced cooperative perception, respectively. 
For each case study, preliminary simulation results were given to demonstrate the superiority of FMs over task-specific models in SoM systems.}
Finally, we compared the existing paradigms for AI-enabled SoM system design and outlined potential future research directions.
\section*{Acknowledgment}
The authors would like to thank Mingran Sun and Zengrui Han for their assistance in the simulation experiments of case study 1 for roadmap 1 and roadmap 2.

\bibliographystyle{IEEEtran}
\small
\bibliography{IEEEabrv, ref}

\begin{thebibliography}{100}
\providecommand{\url}[1]{#1}
\csname url@samestyle\endcsname
\providecommand{\newblock}{\relax}
\providecommand{\bibinfo}[2]{#2}
\providecommand{\BIBentrySTDinterwordspacing}{\spaceskip=0pt\relax}
\providecommand{\BIBentryALTinterwordstretchfactor}{4}
\providecommand{\BIBentryALTinterwordspacing}{\spaceskip=\fontdimen2\font plus
\BIBentryALTinterwordstretchfactor\fontdimen3\font minus \fontdimen4\font\relax}
\providecommand{\BIBforeignlanguage}[2]{{%
\expandafter\ifx\csname l@#1\endcsname\relax
\typeout{** WARNING: IEEEtran.bst: No hyphenation pattern has been}%
\typeout{** loaded for the language `#1'. Using the pattern for}%
\typeout{** the default language instead.}%
\else
\language=\csname l@#1\endcsname
\fi
#2}}
\providecommand{\BIBdecl}{\relax}
\BIBdecl

\bibitem{you2021towards}
X.~You \emph{et~al.}, ``{Towards 6G Wireless Communication Networks: Vision, Enabling Technologies, and New Paradigm Shifts},'' \emph{Sci. China Inf. Sci.}, vol.~64, pp. 1--74, Nov. 2021.

\bibitem{imt2021white}
{IMT-2030 (6G) Promotion Group}, ``{White paper on 6G vision and candidate technologies},'' \emph{China, CAICT}, 2021.

\bibitem{liu2022integrated}
F.~Liu \emph{et~al.}, ``{Integrated Sensing and Communications: Toward Dual-Functional Wireless Networks for 6G and Beyond},'' \emph{{IEEE} J. Select. Areas Commun.}, vol.~40, no.~6, pp. 1728--1767, Jun. 2022.

\bibitem{cheng2023intelligent}
X.~Cheng \emph{et~al.}, ``{Intelligent Multi-Modal Sensing-Communication Integration: Synesthesia of Machines},'' \emph{{IEEE} Commun. Surv. Tutorials}, vol.~26, pp. 258--301, Firstquarter 2024.

\bibitem{liu2023seventy}
F.~Liu \emph{et~al.}, ``{Seventy Years of Radar and Communications: The road from separation to integration},'' \emph{{IEEE} Signal Process Mag.}, vol.~40, no.~5, pp. 106--121, Jul. 2023.

\bibitem{ma2024survey}
Y.~Ma, Z.~Song, Y.~Zhuang, J.~Hao, and I.~King, ``{A Survey on Vision-Language-Action Models for Embodied AI},'' \emph{arXiv preprint arXiv:2405.14093}, 2024.

\bibitem{bai2024multi}
L.~Bai, Z.~Huang, M.~Sun, X.~Cheng, and L.~Cui, ``{Multi-Modal Intelligent Channel Modeling: A New Modeling Paradigm via Synesthesia of Machines},'' \emph{{IEEE} Commun. Surv. Tutorials}, 2025.

\bibitem{cheng2025synthsom}
X.~Cheng \emph{et~al.}, ``{SynthSoM: A Synthetic Intelligent Multi-Modal Sensing-Communication Dataset for Synesthesia of Machines (SoM)},'' \emph{Sci. Data}, vol.~12, no. 819, May 2025.

\bibitem{sun2024multi}
M.~Sun, L.~Bai, Z.~Huang, and X.~Cheng, ``{Multi-Modal Sensing Data Based Real-Time Path Loss Prediction for 6G UAV-to-Ground Communications},'' \emph{IEEE Wireless Commun.}, vol.~13, no.~9, pp. 2462--2466, Sept. 2024.

\bibitem{zhang2024integrated}
H.~Zhang, S.~Gao, X.~Cheng, and L.~Yang, ``{Integrated Sensing and Communications Towards Proactive Beamforming in mmWave V2I via Multi-Modal Feature Fusion (MMFF)},'' \emph{{IEEE} Trans. Wireless Commun.}, vol.~23, pp. 15\,721--15\,735, Nov. 2024.

\bibitem{bommasani2021opportunities}
R.~Bommasani \emph{et~al.}, ``{On the Opportunities and Risks of Foundation Models},'' \emph{arXiv preprint arXiv:2108.07258}, 2021.

\bibitem{zhao2023survey}
W.~X. Zhao \emph{et~al.}, ``{A Survey of Large Language Models},'' \emph{arXiv preprint arXiv:2303.18223}, vol.~1, no.~2, 2023.

\bibitem{liu2024deepseek}
A.~Liu \emph{et~al.}, ``{Deepseek-v3 Technical Report},'' \emph{arXiv preprint arXiv:2412.19437}, 2024.

\bibitem{liang2024foundation}
Y.~Liang \emph{et~al.}, ``{Foundation Models for Time Series Analysis: A Tutorial and Survey},'' in \emph{Proc. ACM SIGKDD Int. Conf. Knowl. Discov. Data Min. (KDD)}, Barcelona, Spain, Aug. 2024, pp. 6555--6565.

\bibitem{bi2023accurate}
K.~Bi, L.~Xie, H.~Zhang, X.~Chen, X.~Gu, and Q.~Tian, ``{Accurate medium-range global weather forecasting with 3D neural networks},'' \emph{Nature}, vol. 619, no. 7970, pp. 533--538, Jul. 2023.

\bibitem{sun2022ringmo}
X.~Sun \emph{et~al.}, ``{RingMo: A Remote Sensing Foundation Model With Masked Image Modeling},'' \emph{{IEEE} Trans. Geosci. Remote Sens.}, vol.~61, pp. 1--22, Jul. 2022.

\bibitem{xu2024large}
M.~Xu \emph{et~al.}, ``{When large language model agents meet 6G networks: Perception, grounding, and alignment},'' \emph{IEEE Wireless Commun.}, vol.~31, no.~6, pp. 63--71, Dec. 2024.

\bibitem{qin2025generative}
X.~Qin \emph{et~al.}, ``{Generative AI Meets Wireless Networking: An Interactive Paradigm for Intent-Driven Communications},'' \emph{IEEE Trans. Cognit. Commun. Networking}, early access 2025.

\bibitem{jiang2024large1}
F.~Jiang \emph{et~al.}, ``{Large Language Model Enhanced Multi-Agent Systems for 6G Communications},'' \emph{IEEE Wireless Commun.}, vol.~31, no.~6, pp. 48--55, Dec. 2024.

\bibitem{zhou2024large}
H.~Zhou \emph{et~al.}, ``{Large Language Model (LLM) for Telecommunications: A Comprehensive Survey on Principles, Key Techniques, and Opportunities},'' \emph{{IEEE} Commun. Surv. Tutorials}, early access 2024.

\bibitem{long2024llms}
L.~Long \emph{et~al.}, ``{On LLMs-Driven Synthetic Data Generation, Curation, and Evaluation: A Survey},'' \emph{arXiv preprint arXiv:2406.15126}, 2024.

\bibitem{liang2024survey}
Z.~Liang \emph{et~al.}, ``{A Survey of Multimodel Large Language Models},'' in \emph{Proc. Int. Conf. Comput. Artif. Intell. Control Eng. (CAICE)}, Xi'an, China, Jan. 2024, pp. 405--409.

\bibitem{vaswani2017attention}
A.~Vaswani \emph{et~al.}, ``{Attention Is All You Need},'' in \emph{Adv. Neural Inf. Process. Syst. (NeurIPS)}, vol.~30, Long Beach, CA, USA, 2017, pp. 6000--6010.

\bibitem{pan2024unifying}
S.~Pan, L.~Luo, Y.~Wang, C.~Chen, J.~Wang, and X.~Wu, ``{Unifying Large Language Models and Knowledge Graphs: A Roadmap},'' \emph{IEEE Trans. Knowl. Data Eng.}, vol.~36, no.~7, pp. 3580--3599, Jul. 2024.

\bibitem{kaplan2020scaling}
J.~Kaplan \emph{et~al.}, ``{Scaling laws for neural language models},'' \emph{arXiv preprint arXiv:2001.08361}, 2020.

\bibitem{ahn2024large}
J.~Ahn, R.~Verma, R.~Lou, D.~Liu, R.~Zhang, and W.~Yin, ``{Large Language Models for Mathematical Reasoning: Progresses and Challenges},'' \emph{arXiv preprint arXiv:2402.00157}, 2024.

\bibitem{guo2023indeed}
T.~Guo \emph{et~al.}, ``{What Indeed Can GPT Models Do In Chemistry? A Comprehensive Benchmark On Eight Tasks},'' \emph{arXiv preprint arXiv:2305.18365}, 2023.

\bibitem{zhang2025scientific}
Q.~Zhang \emph{et~al.}, ``{Scientific Large Language Models: A Survey on Biological \& Chemical Domains},'' \emph{ACM Comput. Surv.}, vol.~57, no.~6, pp. 1--38, Feb. 2025.

\bibitem{hou2024large}
X.~Hou \emph{et~al.}, ``{Large Language Models for Software Engineering: A Systematic Literature Review},'' \emph{ACM Trans. Software Eng. Methodol.}, vol.~33, no.~8, pp. 1--79, Dec. 2024.

\bibitem{han2024parameter}
Z.~Han, C.~Gao, J.~Liu, J.~Zhang, and S.~Q. Zhang, ``{Parameter-Efficient Fine-Tuning for Large Models: A Comprehensive Survey},'' \emph{arXiv preprint arXiv:2403.14608}, 2024.

\bibitem{sahoo2024systematic}
P.~Sahoo, A.~K. Singh, S.~Saha, V.~Jain, S.~Mondal, and A.~Chadha, ``{A Systematic Survey of Prompt Engineering in Large Language Models: Techniques and Applications},'' \emph{arXiv preprint arXiv:2402.07927}, 2024.

\bibitem{zhang2023instruction}
S.~Zhang \emph{et~al.}, ``{Instruction Tuning for Large Language Models: A Survey},'' \emph{arXiv preprint arXiv:2308.10792}, 2023.

\bibitem{shen2024llm}
Z.~Shen, ``{LLM With Tools: A Survey},'' \emph{arXiv preprint arXiv:2409.18807}, 2024.

\bibitem{goyal2024systematic}
M.~Goyal and Q.~H. Mahmoud, ``{A Systematic Review of Synthetic Data Generation Techniques Using Generative AI},'' \emph{Electronics}, vol.~13, no.~17, p. 3509, Sep. 2024.

\bibitem{li2023synthetic}
Z.~Li, H.~Zhu, Z.~Lu, and M.~Yin, ``{Synthetic Data Generation with Large Language Models for Text Classification: Potential and Limitations},'' \emph{arXiv preprint arXiv:2310.07849}, 2023.

\bibitem{abdullin2024synthetic}
Y.~Abdullin, D.~Molla-Aliod, B.~Ofoghi, J.~Yearwood, and Q.~Li, ``{Synthetic Dialogue Dataset Generation using LLM Agents},'' \emph{arXiv preprint arXiv:2401.17461}, 2024.

\bibitem{whitehouse2023llm}
C.~Whitehouse, M.~Choudhury, and A.~F. Aji, ``{LLM-powered Data Augmentation for Enhanced Cross-lingual Performance},'' \emph{arXiv preprint arXiv:2305.14288}, 2023.

\bibitem{kang2024synthetic}
A.~Kang, J.~Y. Chen, Z.~Lee-Youngzie, and S.~Fu, ``{Synthetic Data Generation with LLM for Improved Depression Prediction},'' \emph{arXiv preprint arXiv:2411.17672}, 2024.

\bibitem{tang2023does}
R.~Tang, X.~Han, X.~Jiang, and X.~Hu, ``{Does Synthetic Data Generation of LLMs Help Clinical Text Mining?}'' \emph{arXiv preprint arXiv:2303.04360}, 2023.

\bibitem{he2024zero}
L.~He, H.~Zhang, J.~Liu, K.~Sun, and Q.~Zhang, ``{Zero-Shot Relation Triplet Extraction via Knowledge-Driven LLM Synthetic Data Generation},'' in \emph{Int. Conf. Intell. Comput. (ICIC)}.\hskip 1em plus 0.5em minus 0.4em\relax Springer, 2024, pp. 329--340.

\bibitem{zhou2025lawgpt}
Z.~Zhou \emph{et~al.}, ``{LawGPT: Knowledge-Guided Data Generation and Its Application to Legal LLM},'' \emph{arXiv preprint arXiv:2502.06572}, 2025.

\bibitem{kumichev2024medsyn}
G.~Kumichev \emph{et~al.}, ``{MedSyn: LLM-based Synthetic Medical Text Generation Framework},'' in \emph{Joint Eur. Conf. Mach. Learn. Knowl. Discovery Databases (ECML PKDD)}.\hskip 1em plus 0.5em minus 0.4em\relax Vilnius,Lithuania: Springer, Sep. 2024, pp. 215--230.

\bibitem{fedoseev24constraint}
T.~Fedoseev, D.~I. Dimitrov, T.~Gehr, and M.~Vechev, ``{Constraint-Based Synthetic Data Generation for LLM Mathematical Reasoning},'' in \emph{Workshop on Math. Reasoning AI at NeurIPS'24}, Vancouver, Canada, Dec. 2024.

\bibitem{yang2024enhancing}
D.~Yang, N.~Monaikul, A.~Ding, B.~Tan, K.~Mosaliganti, and G.~Iyengar, ``{Enhancing Table Representations with LLM-powered Synthetic Data Generation},'' \emph{arXiv preprint arXiv:2411.03356}, 2024.

\bibitem{qin2024diffusiongpt}
J.~Qin \emph{et~al.}, ``{DiffusionGPT: LLM-Driven Text-to-Image Generation System},'' \emph{arXiv preprint arXiv:2401.10061}, 2024.

\bibitem{cao2024medical}
X.~Cao \emph{et~al.}, ``{Medical Video Generation for Disease Progression Simulation},'' \emph{arXiv preprint arXiv:2411.11943}, 2024.

\bibitem{zhou2024geng}
X.~Zhou, Q.~Jia, Y.~Hu, R.~Xie, T.~Huang, and F.~R. Yu, ``{GenG: An LLM-Based Generic Time Series Data Generation Approach for Edge Intelligence via Cross-Domain Collaboration},'' in \emph{Proc. IEEE Conf. Comput. Commun. Workshops (INFOCOM WKSHPS)}.\hskip 1em plus 0.5em minus 0.4em\relax Vancouver, Canada: IEEE, May 2024, pp. 1--6.

\bibitem{wang2024harmonic}
Y.~Wang \emph{et~al.}, ``{HARMONIC: Harnessing LLMs for Tabular Data Synthesis and Privacy Protection},'' \emph{arXiv preprint arXiv:2408.02927}, 2024.

\bibitem{gani2023llm}
H.~Gani, S.~F. Bhat, M.~Naseer, S.~Khan, and P.~Wonka, ``{LLM Blueprint: Enabling Text-to-Image Generation with Complex and Detailed Prompts},'' \emph{arXiv preprint arXiv:2310.10640}, 2023.

\bibitem{zhang2024generative}
R.~Zhang \emph{et~al.}, ``{Generative AI-Enabled Vehicular Networks: Fundamentals, Framework, and Case Study},'' \emph{IEEE Network}, vol.~38, no.~4, pp. 259--267, Jul. 2024.

\bibitem{zhang2024generative1}
R.~Zhang \emph{et~al.}, ``{Generative AI agents with large language model for satellite networks via a mixture of experts transmission},'' \emph{IEEE J. Sel. Areas Commun.}, vol.~42, no.~12, pp. 3581--3596, Dec. 2024.

\bibitem{shao2024wirelessllm}
J.~Shao \emph{et~al.}, ``{WirelessLLM: Empowering Large Language Models Towards Wireless Intelligence},'' \emph{arXiv preprint arXiv:2405.17053}, 2024.

\bibitem{kotaru2023adapting}
M.~Kotaru, ``{Adapting Foundation Models for Information Synthesis of Wireless Communication Specifications},'' \emph{arXiv preprint arXiv:2308.04033}, 2023.

\bibitem{he2024designing}
Z.~He \emph{et~al.}, ``{Designing Network Algorithms via Large Language Models},'' in \emph{Proc. ACM Workshop Hot Topics Networks (HotNets)}.\hskip 1em plus 0.5em minus 0.4em\relax New York, NY, USA: Association for Computing Machinery, 2024, pp. 205--212.

\bibitem{medaranga2024poster}
P.~Medaranga, D.~Shah, S.~V. Kandala, and A.~Varshney, ``{POSTER: Simplifying the Networking of Wireless Embedded Systems using a Large Language Model},'' in \emph{Proc. ACM SIGCOMM Posters Demos}, Sydney, NSW, Australia, Aug. 2024, pp. 78--80.

\bibitem{taheri2025domain}
S.~Taheri, A.~Ihalage, P.~Mishra, S.~Coaker, F.~Muhammad, and H.~Al-Raweshidy, ``{Domain Tailored Large Language Models for Log Mask Prediction in Cellular Network Diagnostics},'' \emph{{IEEE} Trans. Netw. Serv. Manage.}, early access 2025.

\bibitem{zou2024telecomgpt}
H.~Zou \emph{et~al.}, ``{TelecomGPT: A Framework to Build Telecom-Specfic Large Language Models},'' \emph{arXiv preprint arXiv:2407.09424}, 2024.

\bibitem{lin2025empowering}
Y.~Lin \emph{et~al.}, ``{Empowering Large Language Models in Wireless Communication: A Novel Dataset and Fine-Tuning Framework},'' \emph{arXiv preprint arXiv:2501.09631}, 2025.

\bibitem{kan2024mobile}
K.~B. Kan, H.~Mun, G.~Cao, and Y.~Lee, ``{Mobile-LLaMA: Instruction Fine-Tuning Open-Source LLM for Network Analysis in 5G Networks},'' \emph{IEEE Network}, vol.~38, pp. 76--83, Sep. 2024.

\bibitem{gajjar2025oransight}
P.~Gajjar and V.~K. Shah, ``{ORANSight-2.0: Foundational LLMs for O-RAN},'' \emph{arXiv preprint arXiv:2503.05200}, 2025.

\bibitem{xiao2024llm}
Z.~Xiao \emph{et~al.}, ``{LLM Agents as 6G Orchestrator: A Paradigm for Task-Oriented Physical-Layer Automation},'' \emph{arXiv preprint arXiv:2410.03688}, 2024.

\bibitem{zhou2024large1}
H.~Zhou \emph{et~al.}, ``{Large Language Model (LLM)-Enabled In-Context Learning for Wireless Network Optimization: A Case Study of Power Control},'' \emph{arXiv preprint arXiv:2408.00214}, 2024.

\bibitem{abbas2024leveraging}
M.~Abbas, K.~Kar, and T.~Chen, ``{Leveraging Large Language Models for Wireless Symbol Detection via In-Context Learning},'' \emph{arXiv preprint arXiv:2409.00124}, 2024.

\bibitem{hu2024self}
C.~Hu, H.~Zhou, D.~Wu, X.~Chen, J.~Yan, and X.~Liu, ``{Self-Refined Generative Foundation Models for Wireless Traffic Prediction},'' \emph{arXiv preprint arXiv:2408.10390}, 2024.

\bibitem{noh2025adaptive}
H.~Noh, B.~Shim, and H.~J. Yang, ``{Adaptive Resource Allocation Optimization Using Large Language Models in Dynamic Wireless Environments},'' \emph{arXiv preprint arXiv:2502.02287}, 2025.

\bibitem{liu2024llm4cp}
B.~Liu, X.~Liu, S.~Gao, X.~Cheng, and L.~Yang, ``{LLM4CP: Adapting Large Language Models for Channel Prediction},'' \emph{J. Commun. Inf. Networks}, vol.~9, no.~2, pp. 113--125, Jun. 2024.

\bibitem{fan2024csi}
S.~Fan, Z.~Liu, X.~Gu, and H.~Li, ``{CSI-LLM: A Novel Downlink Channel Prediction Method Aligned with LLM Pre-Training},'' \emph{arXiv preprint arXiv:2409.00005}, 2024.

\bibitem{sheng2025beam}
Y.~Sheng, K.~Huang, L.~Liang, P.~Liu, S.~Jin, and G.~Y. Li, ``{Beam Prediction Based on Large Language models},'' \emph{IEEE Wireless Commun. Lett.}, early access 2025.

\bibitem{cui2025exploring}
Y.~Cui, J.~Guo, C.-K. Wen, S.~Jin, and E.~Tong, ``{Exploring the Potential of Large Language Models for Massive MIMO CSI Feedback},'' \emph{arXiv preprint arXiv:2501.10630}, 2025.

\bibitem{xue2025large}
J.~Xue \emph{et~al.}, ``{Large AI Model for Delay-Doppler Domain Channel Prediction in 6G OTFS-Based Vehicular Networks},'' \emph{arXiv preprint arXiv:2503.01116}, 2025.

\bibitem{wu2024netllm}
D.~Wu \emph{et~al.}, ``{Netllm: Adapting Large Language Models for Networking},'' in \emph{Proc. ACM SIGCOMM Conf. (SIGCOMM)}, Sydney, NSW, Australia, Aug. 2024, pp. 661--678.

\bibitem{liu2025llm4wm}
X.~Liu, S.~Gao, B.~Liu, X.~Cheng, and L.~Yang, ``{LLM4WM: Adapting LLM for Wireless Multi-Tasking},'' \emph{arXiv preprint arXiv:2501.12983}, 2025.

\bibitem{zheng2024large}
T.~Zheng and L.~Dai, ``{Large Language Model Enabled Multi-Task Physical Layer Network},'' \emph{arXiv preprint arXiv:2412.20772}, 2024.

\bibitem{qin2021semantic}
Z.~Qin, X.~Tao, J.~Lu, W.~Tong, and G.~Y. Li, ``{Semantic Communications: Principles and Challenges},'' \emph{arXiv preprint arXiv:2201.01389}, 2021.

\bibitem{nam2024language}
H.~Nam, J.~Park, J.~Choi, M.~Bennis, and S.-L. Kim, ``{Language-Oriented Communication with Semantic Coding and Knowledge Distillation for Text-to-Image Generation},'' in \emph{Proc. IEEE Int. Conf. Acoust., Speech Signal Process. (ICASSP)}.\hskip 1em plus 0.5em minus 0.4em\relax Seoul, Korea: IEEE, Apr. 2024, pp. 13\,506--13\,510.

\bibitem{guo2023semantic}
S.~Guo, Y.~Wang, S.~Li, and N.~Saeed, ``{Semantic Importance-Aware Communications Using Pre-Trained Language Models},'' \emph{IEEE Commun. Lett.}, vol.~27, no.~9, pp. 2328--2332, Sep. 2023.

\bibitem{wang2024large}
Z.~Wang \emph{et~al.}, ``{Large Language Model Enabled Semantic Communication Systems},'' \emph{arXiv preprint arXiv:2407.14112}, 2024.

\bibitem{pokhrel2024large}
S.~R. Pokhrel and A.~Walid, ``{On Large Language Model Based Joint Source Channel Coding for Semantic Communication},'' in \emph{Int. Conf. Found. Large Lang. Models (FLLM)}.\hskip 1em plus 0.5em minus 0.4em\relax Dubai, United Arab Emirates: IEEE, 2024, pp. 322--329.

\bibitem{jiang2024semantic}
P.~Jiang, C.-K. Wen, X.~Yi, X.~Li, S.~Jin, and J.~Zhang, ``{Semantic communications using foundation models: Design approaches and open issues},'' \emph{IEEE Wireless Commun.}, vol.~31, no.~3, pp. 76--84, Jun. 2024.

\bibitem{jiang2024large}
F.~Jiang \emph{et~al.}, ``{Large AI Model Empowered Multimodal Semantic Communications},'' \emph{IEEE Commun. Mag.}, Jan. 2025.

\bibitem{du2024generative}
H.~Du \emph{et~al.}, ``{Generative Al-aided Joint Training-free Secure Semantic Communications via Multi-modal Prompts},'' in \emph{Proc. IEEE Int. Conf. Acoust., Speech Signal Process. (ICASSP)}.\hskip 1em plus 0.5em minus 0.4em\relax Seoul, Korea: IEEE, Apr. 2024, pp. 12\,896--12\,900.

\bibitem{chen2024semantic}
W.~Chen \emph{et~al.}, ``{Semantic Communication Based on Large Language Model for Underwater Image Transmission},'' \emph{arXiv preprint arXiv:2408.12616}, 2024.

\bibitem{qiao2024latency}
L.~Qiao, M.~B. Mashhadi, Z.~Gao, C.~H. Foh, P.~Xiao, and M.~Bennis, ``{Latency-Aware Generative Semantic Communications With Pre-Trained Diffusion Models},'' \emph{IEEE Wireless Commun. Lett.}, vol.~13, no.~10, pp. 2652--2656, Oct. 2024.

\bibitem{zhao2024lamosc}
Y.~Zhao, Y.~Yue, S.~Hou, B.~Cheng, and Y.~Huang, ``{LaMoSC: Large Language Model-Driven Semantic Communication System for Visual Transmission},'' \emph{IEEE Trans. Cognit. Commun. Networking}, vol.~10, no.~6, pp. 2005--2018, Dec. 2024.

\bibitem{chen2024personalizing}
Z.~Chen, H.~H. Yang, K.~F.~E. Chong, and T.~Q. Quek, ``{Personalizing Semantic Communication: A Foundation Model Approach},'' in \emph{IEEE Workshop Signal Process. Adv. Wireless Commun. (SPAWC)}.\hskip 1em plus 0.5em minus 0.4em\relax Lucca, Italy: IEEE, Sep. 2024, pp. 846--850.

\bibitem{cheng2023m}
X.~Cheng \emph{et~al.}, ``{M$^3$SC: A Generic Dataset for Mixed Multi-Modal (MMM) Sensing and Communication Integration},'' \emph{China Commun.}, vol.~20, no.~11, pp. 13--29, Nov. 2023.

\bibitem{yang2023uniaudio}
D.~Yang \emph{et~al.}, ``{UniAudio: An Audio Foundation Model Toward Universal Audio Generation},'' \emph{arXiv preprint arXiv:2310.00704}, 2023.

\bibitem{chang2022maskgit}
H.~Chang, H.~Zhang, L.~Jiang, C.~Liu, and W.~T. Freeman, ``{MaskGIT: Masked Generative Image Transformer},'' in \emph{Proc. IEEE Conf. Comput. Vis. Pattern Recognit. (CVPR)}, New Orleans, LA, USA, Jun. 2022, pp. 11\,315--11\,325.

\bibitem{bluethgen2024vision}
C.~Bluethgen \emph{et~al.}, ``{A Vision-Language Foundation Model for The Generation of Realistic Chest X-ray Images},'' \emph{Nat. Biomed. Eng.}, vol.~9, pp. 494--506, Aug. 2024.

\bibitem{wang2025mitigating}
W.~Wang, K.~Wu, Y.~B. Li, D.~Wang, X.~Zhang, and J.~Liu, ``{Mitigating Data Scarcity in Time Series Analysis: A Foundation Model with Series-Symbol Data Generation},'' \emph{arXiv preprint arXiv:2502.15466}, 2025.

\bibitem{wu2024vila}
Y.~Wu \emph{et~al.}, ``{VILA-U: a Unified Foundation Model Integrating Visual Understanding and Generation},'' \emph{arXiv preprint arXiv:2409.04429}, 2024.

\bibitem{wu2024mrgen}
H.~Wu, Z.~Zhao, Y.~Zhang, W.~Xie, and Y.~Wang, ``{MRGen: Diffusion-based Controllable Data Engine for MRI Segmentation towards Unannotated Modalities},'' \emph{arXiv preprint arXiv:2412.04106}, 2024.

\bibitem{yang2025revolutionizing}
Z.~Yang \emph{et~al.}, ``{Revolutionizing wireless networks with self-supervised learning: A pathway to intelligent communications},'' \emph{IEEE Wireless Commun.}, early access 2025.

\bibitem{salihu2020low}
A.~Salihu, S.~Schwarz, A.~Pikrakis, and M.~Rupp, ``{Low-dimensional Representation Learning for Wireless CSI-based Localisation},'' in \emph{Int. Conf. Wireless Mobile Comput. Netw. Commun. (WiMob)}.\hskip 1em plus 0.5em minus 0.4em\relax IEEE, Oct. 2020, pp. 1--6.

\bibitem{ferrand2021triplet}
P.~Ferrand, A.~Decurninge, L.~G. Ordonez, and M.~Guillaud, ``{Triplet-Based Wireless Channel Charting: Architecture and Experiments},'' \emph{IEEE J. Sel. Areas Commun.}, vol.~39, no.~8, pp. 2361--2373, Aug. 2021.

\bibitem{naderializadeh2021contrastive}
N.~Naderializadeh, ``{Contrastive Self-Supervised Learning for Wireless Power Control},'' in \emph{Proc. IEEE Int. Conf. Acoust., Speech Signal Process. (ICASSP)}.\hskip 1em plus 0.5em minus 0.4em\relax IEEE, 2021, pp. 4965--4969.

\bibitem{chafaa2022self}
I.~Chafaa, R.~Negrel, E.~V. Belmega, and M.~Debbah, ``{Self-Supervised Deep Learning for mmWave Beam Steering Exploiting Sub-6 GHz Channels},'' \emph{{IEEE} Trans. Wireless Commun.}, vol.~21, no.~10, pp. 8803--8816, Oct. 2022.

\bibitem{davaslioglu2022self}
K.~Davaslioglu, S.~Bozta{\c{s}}, M.~C. Ertem, Y.~E. Sagduyu, and E.~Ayanoglu, ``{Self-Supervised RF Signal Representation Learning for NextG Signal Classification With Deep Learning},'' \emph{IEEE Wireless Commun. Lett.}, vol.~12, no.~1, pp. 65--69, Jan. 2022.

\bibitem{zhao2025transformer}
R.~Zhao, Y.~Ruan, Y.~Li, T.~Li, R.~Zhang, and P.~Xiao, ``{A Transformer based Self-supervised Learning Framework for Robust Time-frequency Localization in Concurrent Cognitive Scenario},'' \emph{{IEEE} Trans. Wireless Commun.}, 2025.

\bibitem{zhang2023self}
Z.~Zhang, T.~Ji, H.~Shi, C.~Li, Y.~Huang, and L.~Yang, ``{A Self-Supervised Learning-Based Channel Estimation for IRS-Aided Communication Without Ground Truth},'' \emph{{IEEE} Trans. Wireless Commun.}, vol.~22, no.~8, pp. 5446--5460, Aug. 2023.

\bibitem{liu2024leveraging}
Z.~Liu \emph{et~al.}, ``{Leveraging Self-Supervised Learning for MIMO-OFDM Channel Representation and Generation},'' \emph{arXiv preprint arXiv:2407.07702}, 2024.

\bibitem{huangfu2019realistic}
Y.~Huangfu \emph{et~al.}, ``{Realistic Channel Models Pre-training},'' in \emph{IEEE Globecom Workshops (GC Wkshps)}.\hskip 1em plus 0.5em minus 0.4em\relax Hawaii, USA: IEEE, Dec. 2019, pp. 1--6.

\bibitem{alikhani2024large}
S.~Alikhani, G.~Charan, and A.~Alkhateeb, ``{Large Wireless Model (LWM): A Foundation Model for Wireless Channels},'' \emph{arXiv preprint arXiv:2411.08872}, 2024.

\bibitem{liu2024wifo}
B.~Liu, S.~Gao, X.~Liu, X.~Cheng, and L.~Yang, ``{WiFo: Wireless Foundation Model for Channel Prediction},'' \emph{Sci. China Inf. Sci.}, early access 2025.

\bibitem{catak2025bert4mimo}
F.~O. Catak, M.~Kuzlu, and U.~Cali, ``{BERT4MIMO: A Foundation Model using BERT Architecture for Massive MIMO Channel State Information Prediction},'' \emph{arXiv preprint arXiv:2501.01802}, 2025.

\bibitem{guo2025prompt}
J.~Guo, Y.~Cui, C.-K. Wen, and S.~Jin, ``{Prompt-Enabled Large AI Models for CSI Feedback},'' \emph{arXiv preprint arXiv:2501.10629}, 2025.

\bibitem{salihu2024self}
A.~Salihu, M.~Rupp, and S.~Schwarz, ``{Self-Supervised and Invariant Representations for Wireless Localization},'' \emph{{IEEE} Trans. Wireless Commun.}, vol.~23, no.~8, pp. 8281--8296, Aug. 2024.

\bibitem{aboulfotouh2024building}
A.~Aboulfotouh, A.~Eshaghbeigi, and H.~Abou-Zeid, ``{Building 6G Radio Foundation Models with Transformer Architectures},'' \emph{arXiv preprint arXiv:2411.09996}, 2024.

\bibitem{zhao2024finding}
Z.~Zhao, T.~Chen, F.~Meng, H.~Li, X.~Li, and G.~Zhu, ``{Finding the missing data: A bert-inspired approach against package loss in wireless sensing},'' in \emph{Proc. IEEE Conf. Comput. Commun. Workshops (INFOCOM WKSHPS)}.\hskip 1em plus 0.5em minus 0.4em\relax IEEE, 2024, pp. 1--6.

\bibitem{zhao2024mining}
Z.~Zhao, F.~Meng, H.~Li, X.~Li, and G.~Zhu, ``{Mining Limited Data Sufficiently: A BERT-inspired Approach for CSI Time Series Application in Wireless Communication and Sensing},'' \emph{arXiv preprint arXiv:2412.06861}, 2024.

\bibitem{jiang2025mimo}
J.~Jiang, W.~Yu, Y.~Li, Y.~Gao, and S.~Xu, ``{A MIMO Wireless Channel Foundation Model via CIR-CSI Consistency},'' \emph{arXiv preprint arXiv:2502.11965}, 2025.

\bibitem{jiao20246g}
T.~Jiao \emph{et~al.}, ``{6G-Oriented CSI-Based Multi-Modal Pre-Ttaining and Downstream Task Adaptation Paradigm},'' in \emph{Int. Conf. Commun. Workshops (ICC Workshops)}.\hskip 1em plus 0.5em minus 0.4em\relax Denver, CO, USA: IEEE, Jun. 2024, pp. 1389--1394.

\bibitem{liu2024timer}
Y.~Liu, H.~Zhang, C.~Li, X.~Huang, J.~Wang, and M.~Long, ``{Timer: Generative Pre-trained Transformers Are Large Time Series Models},'' \emph{arXiv preprint arXiv:2402.02368}, 2024.

\bibitem{shi2024time}
X.~Shi \emph{et~al.}, ``{Time-MoE: Billion-Scale Time Series Foundation Models with Mixture of Experts},'' \emph{arXiv preprint arXiv:2409.16040}, 2024.

\bibitem{zhang2024scaling}
B.~Zhang, Z.~Liu, C.~Cherry, and O.~Firat, ``{When Scaling Meets LLM Finetuning: The Effect of Data, Model and Finetuning Method},'' \emph{arXiv preprint arXiv:2402.17193}, 2024.

\bibitem{xu2024survey}
X.~Xu \emph{et~al.}, ``{A Survey on Knowledge Distillation of Large Language Models},'' \emph{arXiv preprint arXiv:2402.13116}, 2024.

\bibitem{yin2020addressing}
H.~Yin, H.~Wang, Y.~Liu, and D.~Gesbert, ``{Addressing the Curse of Mobility in Massive MIMO With Prony-Based Angular-Delay Domain Channel Predictions},'' \emph{IEEE J. Sel. Areas Commun.}, vol.~38, no.~12, pp. 2903--2917, Dec. 2020.

\bibitem{huang2024lidar}
Z.~Huang, L.~Bai, M.~Sun, and X.~Cheng, ``{A LiDAR-aided channel model for vehicular intelligent sensing-communication integration},'' \emph{IEEE Trans. Intell. Transp. Syst.}, vol.~25, no.~12, pp. 20\,105--20\,119, Dec. 2024.

\bibitem{huang2024scatterer}
Z.~Huang, L.~Bai, Z.~Han, and X.~Cheng, ``Scatterer recognition for multi-modal intelligent vehicular channel modeling via {Synesthesia} of {Machines},'' \emph{IEEE Wireless Commun.}, early access 2025.

\bibitem{han2025llm4sp}
Z.~Han, L.~Bai, Z.~Huang, and X.~Cheng, ``Llm4sg: Large language models for scatterer generation via synesthesia of machines,'' \emph{arXiv preprint arXiv:2505.17879}, 2025.

\bibitem{radford2019language}
A.~Radford \emph{et~al.}, ``{Language Models are Unsupervised Multitask Learners},'' \emph{OpenAI blog}, vol.~1, no.~8, p.~9, Feb. 2019.

\bibitem{qi2022parameter}
W.~Qi, Y.-P. Ruan, Y.~Zuo, and T.~Li, ``{Parameter-Efficient Tuning on Layer Normalization for Pre-trained Language Models},'' \emph{arXiv preprint arXiv:2211.08682}, 2022.

\bibitem{soltani2019deep}
M.~Soltani, V.~Pourahmadi, A.~Mirzaei, and H.~Sheikhzadeh, ``{Deep Learning-Based Channel Estimation},'' \emph{IEEE Commun. Lett.}, vol.~23, no.~4, pp. 652--655, Apr. 2019.

\bibitem{salihu2022attention}
A.~Salihu, S.~Schwarz, and M.~Rupp, ``{Attention Aided CSI Wireless Localization},'' in \emph{IEEE Workshop Signal Process. Adv. Wireless Commun. (SPAWC)}.\hskip 1em plus 0.5em minus 0.4em\relax Oulu, Finland: IEEE, Jul. 2022, pp. 1--5.

\bibitem{he2016deep}
K.~He, X.~Zhang, S.~Ren, and J.~Sun, ``{Deep Residual Learning for Image Recognition},'' in \emph{Proc. IEEE Conf. Comput. Vis. Pattern Recognit. (CVPR)}, Las Vegas, NV, USA, Jun. 2016, pp. 770--778.

\bibitem{misra2016cross}
I.~Misra, A.~Shrivastava, A.~Gupta, and M.~Hebert, ``{Cross-Stitch Networks for Multi-task Learning},'' in \emph{Proc. IEEE Conf. Comput. Vis. Pattern Recognit. (CVPR)}, Las Vegas, NV, USA, Jun. 2016, pp. 3994--4003.

\bibitem{liu2024datasets}
Y.~Liu, J.~Cao, C.~Liu, K.~Ding, and L.~Jin, ``{Datasets for Large Language Models: A Comprehensive Survey},'' \emph{arXiv preprint arXiv:2402.18041}, 2024.

\bibitem{geiger2013vision}
A.~Geiger, P.~Lenz, C.~Stiller, and R.~Urtasun, ``{Vision Meets Robotics: The Kitti Dataset},'' \emph{Int. J. Rob. Res.}, vol.~32, no.~11, pp. 1231--1237, Nov. 2013.

\bibitem{alkhateeb2023deepsense}
A.~Alkhateeb \emph{et~al.}, ``{DeepSense 6G: A Large-Scale Real-World Multi-Modal Sensing and Communication Dataset},'' \emph{{IEEE} Commun. Mag.}, vol.~61, no.~9, pp. 122--128, Sep. 2023.

\bibitem{alrabeiah2020viwi}
M.~Alrabeiah, A.~Hredzak, Z.~Liu, and A.~Alkhateeb, ``{ViWi: A Deep Learning Dataset Framework for Vision-Aided Wireless Communications},'' in \emph{Proc. IEEE Veh. Technol. Conf. (VTC2020-Spring)}.\hskip 1em plus 0.5em minus 0.4em\relax Antwerp, Belgium: IEEE, May 2020, pp. 1--5.

\bibitem{aiello2022cross}
E.~Aiello, D.~Valsesia, and E.~Magli, ``Cross-modal learning for image-guided point cloud shape completion,'' \emph{Advances in Neural Information Processing Systems}, vol.~35, pp. 37\,349--37\,362, 2022.

\bibitem{zheng2023autofed}
T.~Zheng, A.~Li, Z.~Chen, H.~Wang, and J.~Luo, ``Autofed: Heterogeneity-aware federated multimodal learning for robust autonomous driving,'' in \emph{Proceedings of the 29th annual international conference on mobile computing and networking}, 2023, pp. 1--15.

\bibitem{dosovitskiy2020image}
A.~Dosovitskiy \emph{et~al.}, ``{An Image is Worth 16x16 Words: Transformers for Image Recognition at Scale},'' \emph{arXiv preprint arXiv:2010.11929}, 2020.

\bibitem{wang2024scaling}
L.~Wang, X.~Chen, J.~Zhao, and K.~He, ``{Scaling Proprioceptive-Visual Learning with Heterogeneous Pre-trained Transformers},'' in \emph{Adv. Neural Inf. Process. Syst. (NeurIPS)}, vol.~37, Vancouver, Canada, Dec. 2024, pp. 124\,420--124\,450.

\bibitem{fedus2022switch}
W.~Fedus, B.~Zoph, and N.~Shazeer, ``{Switch Transformers: Scaling to Trillion Parameter Models with Simple and Efficient Sparsity},'' \emph{J. Mach. Learn. Res.}, vol.~23, no. 120, pp. 1--39, Jan. 2022.

\bibitem{jiang2024mixtral}
A.~Q. Jiang \emph{et~al.}, ``{Mixtral of Experts},'' \emph{arXiv preprint arXiv:2401.04088}, 2024.

\bibitem{dai2024deepseekmoe}
D.~Dai \emph{et~al.}, ``{DeepSeekMoE: Towards Ultimate Expert Specialization in Mixture-of-Experts Language Models},'' \emph{arXiv preprint arXiv:2401.06066}, 2024.

\bibitem{shazeer2020glu}
N.~Shazeer, ``{Glu Variants Improve Transformer},'' \emph{arXiv preprint arXiv:2002.05202}, 2020.

\bibitem{touvron2023llama}
H.~Touvron \emph{et~al.}, ``{LLaMA: Open and Efficient Foundation Language Models},'' \emph{arXiv preprint arXiv:2302.13971}, 2023.

\bibitem{dao2022flashattention}
T.~Dao, D.~Fu, S.~Ermon, A.~Rudra, and C.~R{\'e}, ``{FlashAttention: Fast and Memory-Efficient Exact Attention with IO-Awareness},'' in \emph{Adv. Neural Inf. Process. Syst. (NeurIPS)}, vol.~35, Vancouver, Canada, Jun. 2022, pp. 16\,344--16\,359.

\bibitem{ainslie2023gqa}
J.~Ainslie, J.~Lee-Thorp, M.~De~Jong, Y.~Zemlyanskiy, F.~Lebr{\'o}n, and S.~Sanghai, ``{GQA: Training Generalized Multi-Query Transformer Models from Multi-Head Checkpoints},'' \emph{arXiv preprint arXiv:2305.13245}, 2023.

\bibitem{liu2021swin}
Z.~Liu \emph{et~al.}, ``{Swin Transformer: Hierarchical Vision Transformer using Shifted Windows},'' in \emph{Proc. IEEE Conf. Comput. Vis. Pattern Recognit. (CVPR)}, Nashville, TN, USA, Jun. 2021, pp. 10\,012--10\,022.

\bibitem{deng2009imagenet}
J.~Deng, W.~Dong, R.~Socher, L.-J. Li, K.~Li, and L.~Fei-Fei, ``{Imagenet: A Large-Scale Hierarchical Image Database},'' in \emph{Proc. IEEE Conf. Comput. Vis. Pattern Recognit. (CVPR)}.\hskip 1em plus 0.5em minus 0.4em\relax Miami, FL, USA: IEEE, Jun. 2009, pp. 248--255.

\bibitem{guo2024deep}
J.~Guo, T.~Chen, S.~Jin, G.~Y. Li, X.~Wang, and X.~Hou, ``{Deep Learning for Joint Channel Estimation and Feedback in Massive MIMO Systems},'' \emph{Digital Commun. Networks}, vol.~10, no.~1, pp. 83--93, Feb. 2024.

\bibitem{he2022masked}
K.~He, X.~Chen, S.~Xie, Y.~Li, P.~Doll{\'a}r, and R.~Girshick, ``{Masked Autoencoders Are Scalable Vision Learners},'' in \emph{Proc. IEEE Conf. Comput. Vis. Pattern Recognit. (CVPR)}, New Orleans, LA, USA, Jun. 2022, pp. 16\,000--16\,009.

\bibitem{chen2020simple}
T.~Chen, S.~Kornblith, M.~Norouzi, and G.~Hinton, ``{A Simple Framework for Contrastive Learning of Visual Representations},'' in \emph{Int. Conf. Mach. Learn. (ICML)}, Jul. 2020, pp. 1597--1607.

\bibitem{zhang2024cloud}
J.~Zhang, Z.~Wei, B.~Liu, X.~Wang, Y.~Yu, and R.~Zhang, ``{Cloud-Edge-Terminal Collaborative AIGC for Autonomous Driving},'' \emph{IEEE Wireless Commun.}, vol.~31, no.~4, pp. 40--47, 2024.

\bibitem{yaman2024luvira}
I.~Yaman \emph{et~al.}, ``{The LuViRA Dataset: Synchronized Vision, Radio, and Audio Sensors for Indoor Localization},'' in \emph{Proc. Int. Conf. Robot. Automat. (ICRA)}.\hskip 1em plus 0.5em minus 0.4em\relax Yokohama, Japan: IEEE, May 2024, pp. 11\,920--11\,926.

\bibitem{dichasus2021}
F.~Euchner, M.~Gauger, S.~D\"orner, and S.~ten Brink, ``{A Distributed Massive MIMO Channel Sounder for "Big CSI Data"-driven Machine Learning},'' in \emph{Proc. Int. ITG Workshop Smart Antennas (WSA)}, Eurecom, France, Nov. 2021, pp. 1--6.

\bibitem{houlsby2019parameter}
N.~Houlsby \emph{et~al.}, ``{Parameter-Efficient Transfer Learning for NLP},'' in \emph{Proc. Int. Conf. Mach. Learn. (ICML)}, Los Angeles, CA, USA, Jul. 2019, pp. 2790--2799.

\end{thebibliography}

\end{document}